\documentclass[a4paper,fleqn]{cas-dc}

\renewcommand{\dfrwsconference}{Digital Forensics Research Conference (\href{https://dfrws.org/presentation/started-off-local-now-were-in-the-cloud-forensic-examination-of-the-amazon-echo-show-15-smart-display/}{DFRWS USA 2024})}

\let\printorcid\relax

\usepackage[dvipsnames]{xcolor}

\usepackage{inconsolata}

\usepackage{tabularx}
\usepackage{makecell}
\usepackage{booktabs}
\usepackage{multicol}
\usepackage{multirow}
\usepackage{threeparttable}

\graphicspath{{figures/}}
\usepackage{tikz}
\usepackage{subcaption}

\usepackage{mdframed}

\usepackage{listings}
\usepackage{lstautogobble}

\usepackage[absolute]{textpos}
\usepackage{calc}

\usepackage{paralist}
\usepackage[inline]{enumitem}

\usepackage[authoryear]{natbib}

\usepackage{cleveref}

\usepackage{acro}[=v3]
\acsetup{single=true}

\usepackage{hyphenat}

\usepackage{xurl}

\usepackage[colorlinks]{hyperref}
\hypersetup{
    linkcolor=RoyalBlue,
    citecolor=RoyalBlue,
    urlcolor=RoyalBlue
}

\makeatletter
\newcommand\notsotiny{\@setfontsize\notsotiny{7.01}{7}}
\makeatother

\crefformat{section}{\S#2#1#3} 
\crefformat{subsection}{\S#2#1#3}
\crefformat{subsubsection}{\S#2#1#3}

\newlist{questions}{enumerate}{2}
\setlist[questions,1]{label=RQ\arabic*.,ref=RQ\arabic*}
\setlist[questions,2]{label=(\alph*),ref=\thequestionsi(\alph*)}

\setlist[itemize]{topsep=0.2em,
                  itemsep=-0.2em,
                  leftmargin=1.4em,
}

\setlist[enumerate]{topsep=0.2em,
                  itemsep=-0.2em,
                  leftmargin=1.4em,
}

\newlist{filepaths}{description}{1}
\setlist[filepaths,1]{font=\mdseries,
                      topsep=0.2em,
                      itemsep=-0.1em,
                      labelsep=0.8em,
                      leftmargin=1.4em,
                      style=unboxed,
}

\newcommand*\emptycirc[1][1ex]{\tikz\draw (0,0) circle (#1);} 
\newcommand*\halfcirc[1][1ex]{%
  \begin{tikzpicture}
  \draw[fill] (0,0)-- (90:#1) arc (90:270:#1) -- cycle ;
  \draw (0,0) circle (#1);
  \end{tikzpicture}}
\newcommand*\fullcirc[1][1ex]{\tikz\fill (0,0) circle (#1);} 

\newcommand{\colorcirc}[1]{
  {\tikz\draw[#1,fill=#1] (0,0) circle (.8ex);}
}

\newcommand{\ballnumber}[1]{\tikz[baseline=(myanchor.base)] \node[circle,fill=.,inner sep=0.9pt] (myanchor) {\color{-.}\bfseries\footnotesize #1};}

\definecolor{pcb_red}{HTML}{ff0000}
\definecolor{pcb_purple}{HTML}{cc00ff}
\definecolor{pcb_green}{HTML}{66ff00}
\definecolor{pcb_blue}{HTML}{00ccff}
\definecolor{pcb_orange}{HTML}{ff6600}
\definecolor{pcb_yellow}{HTML}{ffcc00}

\DeclareAcronym{FAU}{short=FAU, long=Friedrich-Alexander-Universit{\"a}t Erlangen-N{\"u}rnberg}

\DeclareAcronym{API}{
    short=API,
    long=application programming interface,
    short-plural-form=APIs,
    long-plural-form=application programming interfaces
}

\DeclareAcronym{ALS}{
    short=ALS,
    long=ambient light sensor,
    short-plural-form=ALSs,
    long-plural-form=ambient light sensors
}

\DeclareAcronym{CPU}{
    short=CPU,
    long=central processing unit,
    short-plural-form=CPUs,
    long-plural-form=central processing units
}

\DeclareAcronym{GPU}{
    short=GPU,
    long=graphics processing unit,
    short-plural-form=GPUs,
    long-plural-form=graphics processing units
}

\DeclareAcronym{ADB}{
    short=ADB,
    long=Android Debug Bridge
}

\DeclareAcronym{AVD}{
    short=AVD,
    long=Android Virtual Device,
    short-plural-form=AVDs,
    long-plural-form=Android Virtual Devices
}

\DeclareAcronym{AVS}{
    short=AVS,
    long=Alexa Voice Service,
    short-plural-form=AVSs,
    long-plural-form=Alexa Voice Services
}

\DeclareAcronym{IVA}{
    short=IVA,
    long=intelligent virtual assistant,
    short-plural-form=IVAs,
    long-plural-form=intelligent virtual assistants
}

\DeclareAcronym{IPA}{
    short=IPA,
    long=Intelligent Personal Assistant,
    short-plural-form=IPAs,
    long-plural-form=Intelligent Personal Assistants
}

\DeclareAcronym{AWS}{
    short=AWS,
    long=Amazon Web Services,
}

\DeclareAcronym{WLAN}{
    short=WLAN,
    long=wireless local area network,
}

\DeclareAcronym{JSON}{
    short=JSON,
    long=JavaScript Object Notation,
}

\DeclareAcronym{DIAL}{
    short=DIAL,
    long=Discovery And Launch,
}

\DeclareAcronym{NSCA}{
    short=NSCA,
    long=Nagios Service Check Acceptor,
}

\DeclareAcronym{HDMI}{
    short=HDMI,
    long=High Definition Multimedia Interface,
}

\DeclareAcronym{TLS}{
    short=TLS,
    long=Transport Layer Security,
}

\DeclareAcronym{SSDP}{
    short=SSDP,
    long=Simple Service Discovery Protocol,
}

\DeclareAcronym{SSDP:NT}{
    short=NT,
    long=Notification Type,
    long-plural-form=notification types,
}

\DeclareAcronym{UPnP}{
    short=UPnP,
    long=Universal Plug and Play,
}

\DeclareAcronym{VM}{
    short=VM,
    long=virtual machine,
    short-plural-form=VMs,
    long-plural-form=virtual machines
}

\DeclareAcronym{APK}{
    short=APK,
    long=Android Package,
    short-plural-form=APKs,
    long-plural-form=Android Packages
}

\DeclareAcronym{BLE}{
    short=BLE,
    long=Bluetooth Low Energy,
}

\DeclareAcronym{MITM}{
    short=MITM,
    long=man-in-the-middle,
}

\DeclareAcronym{PCB}{
    short=PCB,
    long=printed circuit board,
    short-plural-form=PCBs,
    long-plural-form=printed circuit boards
}

\DeclareAcronym{EXIF}{
    short=EXIF,
    long=Exchangeable Image File Format,
}

\DeclareAcronym{SoC}{
    short=SoC,
    long=system on chip,
    short-plural-form=SoCs,
    long-plural-form=systems on chip,
}

\DeclareAcronym{PCB:FFC}{
    short=FFC,
    long=flexible flat cable,
    short-plural-form=FFCs,
    long-plural-form=flexible flat cables,
}

\DeclareAcronym{BLOB}{
    short=BLOB,
    long=binary large object,
    short-plural-form=BLOBs,
    long-plural-form=binary large objects,
}

\DeclareAcronym{OS}{
    short=OS,
    long=operating system,
    short-plural-form=OSs,
    long-plural-form=operating systems,
}

\DeclareAcronym{CLI}{
    short=CLI,
    long=command line interface,
    short-plural-form=CLIs,
    long-plural-form=command line interfaces,
}

\DeclareAcronym{SSID}{
    short=SSID,
    long=Service Set Identifier,
    short-plural-form=SSIDs,
    long-plural-form=Service Set Identifiers,
}

\DeclareAcronym{UUID}{
    short=UUID,
    long=universally unique identifier,
    short-plural-form=UUIDs,
    long-plural-form=universally unique identifiers,
}

\DeclareAcronym{UART}{
    short=UART,
    long=universal asynchronous receiver-transmitter,
}

\DeclareAcronym{JTAG}{
    short=JTAG,
    long=Joint Test Action Group,
}

\DeclareAcronym{ISP}{
    short=ISP,
    long=in-system programming,
}

\DeclareAcronym{MMC}{
    short=MMC,
    long=MultiMediaCard,
    short-plural-form=MMCs,
    long-plural-form=MultiMediaCards,
}

\DeclareAcronym{eMMC}{
    short=eMMC,
    long=embedded MultiMediaCard,
    short-plural-form=eMMCs,
    long-plural-form=embedded MultiMediaCards,
}

\DeclareAcronym{FCC}{
    short=FCC,
    long=Federal Communications Commission,
}

\DeclareAcronym{CIFT}{
    short=CIFT,
    long=Cloud-based IoT Forensic Toolkit,
}

\DeclareAcronym{BGA}{
    short=BGA,
    long=Ball Grid Array,
}

\DeclareAcronym{MLB}{
    short=MLB,
    long=main logic board,
    short-plural-form=MLBs,
    long-plural-form=main logic boards,
}

\DeclareAcronym{CBC}{
    short=CBC,
    long=cipher-block chaining,
}

\DeclareAcronym{TNR}{
    short=TNR,
    long=temporal noise reduction,
}

\DeclareAcronym{TP}{
    short=TP,
    long=test point,
    short-plural-form=TPs,
    long-plural-form=test points,
}
\begin{document}
\let\WriteBookmarks\relax
\def\floatpagepagefraction{1}
\def\textpagefraction{.001}


\title[mode=title]{\emph{Started Off Local, Now We're in the Cloud}: Forensic Examination of the Amazon Echo Show 15 Smart Display}

\shorttitle{\emph{Started Off Local, Now We're in the Cloud}: Forensic Examination of the Amazon Echo Show 15 Smart Display}

\shortauthors{Crasselt \& Pugliese}  


\author[1]{Jona Crasselt}

\ead{jona.crasselt@fau.de}

\affiliation[1]{organization={Friedrich-Alexander-Universit{\"a}t Erlangen-N{\"u}rnberg (FAU)}, country={Germany}}

\cormark[1]
\cortext[1]{corresponding author}

\credit{
Conceptualization, %
Data curation, %
Investigation, %
Methodology, %
Visualization, %
Software, %
Validation, %
Writing --- original draft, %
Writing --- review \& editing%
}


\author[1]{Gaston Pugliese}

\credit{
Conceptualization, %
Investigation, %
Methodology, %
Visualization, %
Resources, %
Project administration, %
Funding acquisition, %
Writing --- review \& editing%
}


\begin{abstract}
\emph{Amazon Echo} is one of the most popular 
product families of smart speakers and displays. %
Considering their growing presence in modern households 
as well as the digital traces 
associated with residents' interactions with these devices, 
analyses of Echo products are likely to become more common 
for forensic investigators at ``smart home'' crime scenes. %
With this in mind, we present the first forensic examination 
of the \emph{Echo Show~15}, Amazon's largest smart display 
running on \emph{Fire OS} and the first Echo device 
with \emph{Visual ID}, a face recognition feature. %
We unveil a non-invasive method for accessing 
the unencrypted file system of the Echo Show~15
based on an undocumented pinout 
for the \acs{eMMC} interface 
which we discovered on the \acl{MLB}. %
On the device, we identify various local usage artifacts, 
such as searched products, streamed movies, 
visited websites, metadata of photos and videos 
as well as logged events of Visual ID 
about movements and users detected by the built-in camera. %
Furthermore, we utilize an insecurely stored token on the Echo Show~15 
to obtain access to remote user artifacts in Amazon's cloud, 
including Alexa voice requests, calendars, contacts, conversations, 
photos, and videos. %
In this regard, we also identify new Amazon \acsp{API} 
through network traffic analysis of two companion apps, 
namely \emph{Alexa} and \emph{Photos}. %
Overall, in terms of practical relevance, 
our findings demonstrate a non-destructive way of data acquisition 
for Echo Show 15 devices 
as well as how to lift the scope of forensic traces 
from local artifacts on the device 
to remote artifacts stored in the cloud. 
\end{abstract}

\begin{textblock*}{11cm}(7.7cm,3.7cm)
  \begin{minipage}[t][1cm][t]{11cm}
\begin{flushright}
\textcolor{black!50}{
   Accepted for publication at \textbf{DFRWS USA 2024}.\\%
   \href{https://dfrws.org/wp-content/uploads/2024/07/dfrws-usa-2024-echo-show-15.pdf}{Original PDF} available at \href{https://dfrws.org/presentation/started-off-local-now-were-in-the-cloud-forensic-examination-of-the-amazon-echo-show-15-smart-display/}{conference website}.\\%
   \textsc{Bib}\TeX{} entry available at \href{https://github.com/jcrasselt/amazon-echo-show-15\#paper}{GitHub}.%
}
\end{flushright}
  \end{minipage}
\end{textblock*}

\maketitle


\section{Introduction}
\label{introduction}

Smart home appliances often provide a low-threshold entry
into the \emph{smartification} of domestic living environments. %
Especially smart speakers with their ability to play music, 
answer questions via \acp{IVA}, or control other connected devices, 
are a strikingly popular category of smart home products \citep{statista2022homesgettingsmarter}. 
Due to their display-less design, 
the interaction with smart speakers is primarily driven 
by users' voice requests which are answered by audio responses. %
For certain actions, or for receiving visual responses, 
however, a smartphone---usually in combination with a 
companion app---is required. %
This interaction gap was closed by 
the product category of \emph{smart displays} 
whose built-in screens and graphical user interfaces 
enable direct touch control in addition. %

Amazon's smart device family \textit{Echo} 
was by far the most popular brand 
in Germany and the U.S. in 2022 
\citep{
    statista2021popularsmartspeaker, 
    statista2023amazondominatessmartspeaker}. %
The Echo series features the \ac{IVA} \emph{Alexa} 
and includes mainly smart speakers and displays 
but also earbuds and glasses. %
Since Amazon has launched its Echo product family in 2014, 
new devices have been released regularly. %
In 2021, a new wall- or stand-mounted smart display model was introduced  
and marketed as a dashboard for family organization, 
home automation, and streaming: %
The \emph{Echo Show~15} 
is Amazon's largest smart display at the time of writing 
and the first model that shipped with \emph{Visual ID}, 
a user recognition feature enabled by the built-in camera.

Due to their popularity and increasing presence in smart home environments 
as well as 
the variety of traces resulting from active user interactions 
and passive observations of the environment via microphone and camera, 
Echo devices are interesting objects of investigation
from both a privacy and forensic perspective
\citep{lau2018alexa, bouchaud2018iot}. %
Prior work on previous Echo devices showed that 
user-related data is only partly stored locally, 
whereas the Amazon cloud is a more comprehensive source 
\citep{
    chung2017alexaecosystem, 
    krueger2020using, 
    youn2021echoshow}, %
which highlights the importance of obtaining both local and remote artifacts 
during forensic investigations. 

As for IoT devices \citep{gomez2021developing}, however, 
and despite their similarities with desktop or mobile devices, 
the forensic analysis of non-conventional devices, 
like those of the Echo series, 
can be challenging and time-consuming. %
Practically speaking, unique hardware, custom firmware, and security measures 
can hinder analyses in terms of success and economic efficiency. %
Using a hardware debug interface (e.g., UART, JTAG) 
or removing the flash memory chip (``\emph{chip-off}'')
is often the only option for data gathering; 
at least if an interface is available and the data is not encrypted. %
Besides, if no exploitable vulnerabilities exist, 
and if both the user and the vendor are unwilling to cooperate, 
obtaining credentials for bypassing authentication or encryption mechanisms  
is another obstacle for forensic investigators. %
Likewise, since acquisition methods vary in their invasiveness, 
it is crucial to assess and minimize 
the risk of data damage and loss 
to ensure \emph{forensic soundness} \citep{casey2007does}. %

In light of the fact that smart devices become 
``\emph{invisible witnesses}'' \citep{urquhart2022policing} 
at crime scenes in smart homes, 
this paper presents the first forensic examination of the Echo Show~15 
to close a device-specific research gap in the forensic literature. %
As is typical for a device that has not been thoroughly studied yet, 
we began our research by asking the fundamental question: %
\emph{Which local and remote artifacts of forensic relevance 
can be acquired from the Amazon Echo Show~15, and how?}

\paragraph{Contributions.} %

As we are not aware of any prior work on the Echo Show~15, 
our main contributions are as follows: %

\begin{itemize}
\item To the best of our knowledge, we are the first to 
examine Amazon's smart display Echo Show~15 forensically. %
\item We identified an undocumented \acs{eMMC} pinout on the \ac{MLB}, 
resulting in non-invasive access to the unencrypted file system of the Echo Show~15. %
\item We discovered that the Echo Show~15 logs events of detected movements 
and recognized users locally. %
Besides identifying device-specific artifacts, %
we also confirmed the existence of artifacts 
known from prior Echo devices. %
\item We decrypted a token from a local database of the Echo Show~15 
for accessing remote artifacts via cloud \acsp{API}. %
\item We encountered additional Amazon \acs{API} endpoints 
by analyzing the network traffic of two companion apps, 
namely \emph{Alexa} and \emph{Photos}, 
resulting in new and alternative sources for obtaining remote artifacts. %
\end{itemize}

\paragraph{Artifacts.} %

Our paper artifacts are open source and available at 
\href{https://github.com/jcrasselt/amazon-echo-show-15}{github.com/jcrasselt/amazon-echo-show-15}, 
including curated overviews on both local and remote artifacts of the Echo Show~15
as well as scripts to ease data acquisition.

\paragraph{Outline.} %

Initially, we provide background information on the Echo Show~15 and related work (\Cref{background}). %
Our examination results of the Echo Show~15 are divided into three sections,  
namely hardware findings (\Cref{hardware-examination}), 
file system findings (\Cref{filesystem-artifacts}), 
and companion app and cloud findings (\Cref{cloud-app}). %
Finally, we discuss our results (\Cref{discussion}), 
and conclude the paper (\Cref{conclusion}).

\section{Background}
\label{background}

In this section, we first provide technical and contextual information 
about the Echo Show~15 smart display by addressing its hardware, software, user accounts, 
as well as its cloud connectivity and companion apps. %
Afterward, we give an overview on related work in the forensic literature. %

\subsection{Amazon Echo Show 15}
\label{background:echo-show-15}

\begin{figure}[thb]
  \centering
  \begin{subfigure}[b]{\linewidth}
    \centering
    \includegraphics[width=0.9\linewidth]{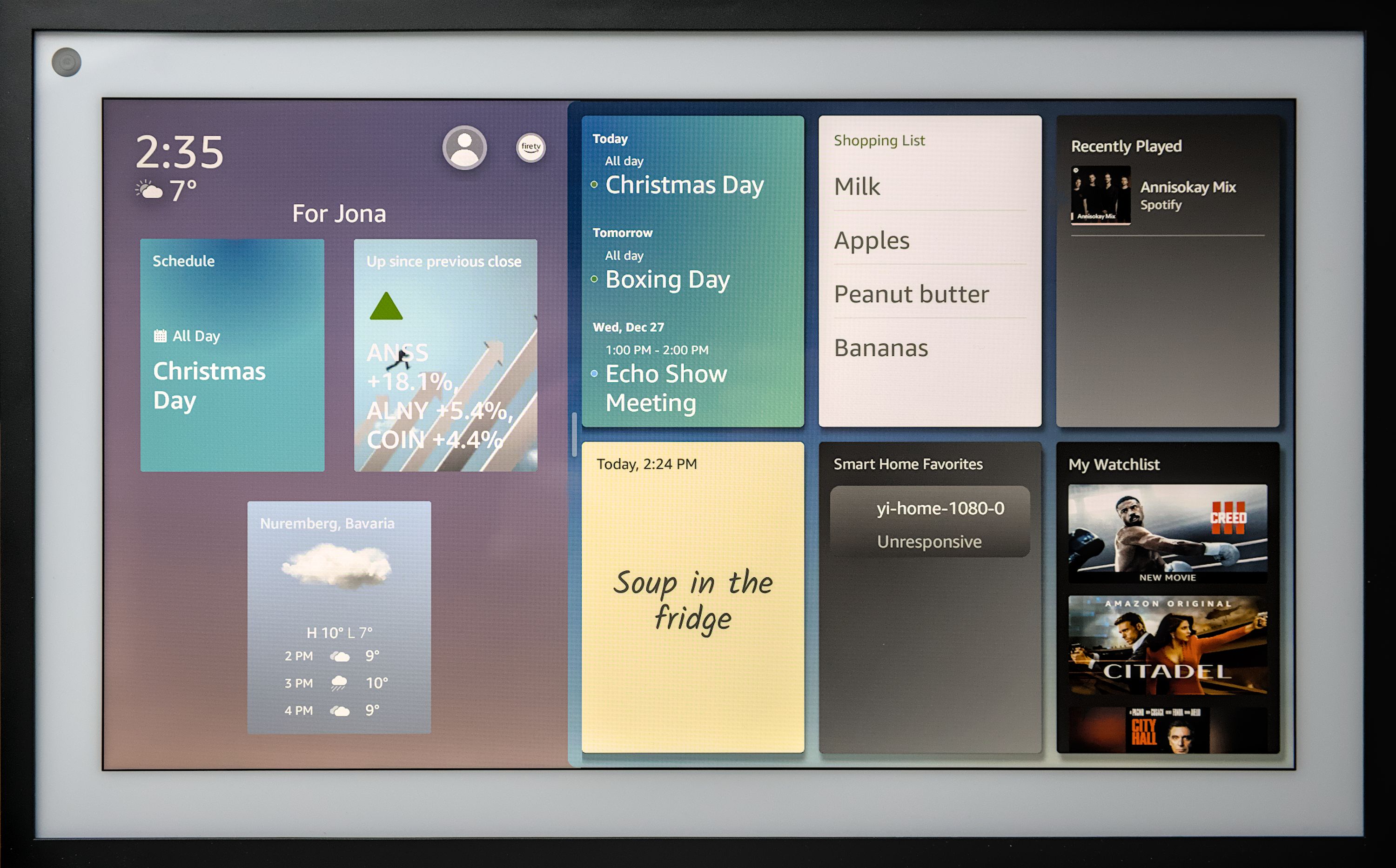}
    \caption{Display showing main view with widgets.}
    \label{fig:background:echo-show-15:display}
  \end{subfigure}

  \vspace*{0.5em}

  \begin{subfigure}[b]{\linewidth}
    \centering
    \includegraphics[width=0.9\linewidth]{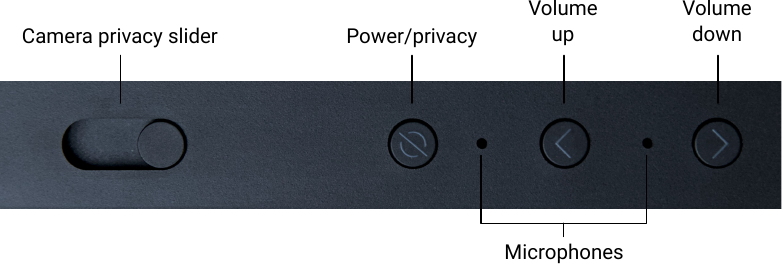}
    \caption{Buttons and slider on top of casing frame.}
    \label{fig:background:echo-show-15:buttons}
  \end{subfigure}

  \caption{Echo Show 15 -- User interface and analog controls.}
  \label{fig:background:echo-show-15}
\end{figure}

The Echo Show~15 is Amazon's first wall- and stand-mountable 
smart display and marketed for family organization and 
entertainment purposes, as well as for serving as a digital photo frame 
and smart home hub \citep{amazon2022devicespecifications}. %
Since its release in 2021, the Echo Show~15 has the largest screen 
among all models and device generations within the Echo Show product family, 
as the other smart displays were primarily designed to be placed on a counter 
(e.g., Echo Show~5, Echo Show~8, or Echo Show~10). %

\paragraph{Hardware.} %

The Echo Show~15 (cf. \Cref{fig:background:echo-show-15:display}) 
has a housing dimension of 402x252x35mm (WxHxD) 
and features a 15.6-inch Full HD touchscreen (1080p) 
supporting both landscape and portrait mode 
\citep{amazon2022devicespecifications}. %
The device specifications 
further disclose an ``Amlogic PopcornA (Pop1)'' \ac{SoC} 
with four ``Arm Cortex-A73'' and four ``Arm Cortex-A53'' \acsp{CPU}, 
an ``Arm Mali-G52 MP8(8EE)'' \acs{GPU}, as well as 
3GB of RAM and 16GB of flash memory (\acs{eMMC}). %
The smart display has a 5~MP front-facing camera, 
a 6-mic array, an \acl{ALS} (\acs{ALS}), and two speakers. %
Wireless communication is supported 
via Wi-Fi 802.11 (a/b/g/n/ac) and Bluetooth~5.0. %
Apart from a power supply jack, 
there is only a Micro-USB port 
to connect a USB-to-Ethernet adapter 
for wired Internet access. %
The casing frame contains a physical slider to cover the camera
and three analog buttons (cf. \Cref{fig:background:echo-show-15:buttons}); 
two buttons to control the speaker volume, and one button to turn the 
device on and off and also to activate a privacy mode 
which prevents the virtual assistant \emph{Alexa} 
from listening to voice prompts. %

\paragraph{Software.} %

Like Amazon's \emph{Fire~TV} and \emph{Fire~Tablet} 
devices \citep{amazon2017fireosoverview}, 
the Echo Show 15 runs \emph{Fire OS}~7 
which is based on Android~9 (Level~28). %
The scope of functions, however, has been
reduced for Echo Show devices, as, for example, 
sideloading apps is no longer possible 
\citep{iobroker2023echoshow15apks}. %
The Echo Show~15 can be used to, inter alia, engage with Alexa via voice commands, 
manage calendars or to-do and shopping lists, 
stream music and movies, make video calls, or browse the Web using the \emph{Silk} browser. %
Also, users can be distinguished by their voice and face 
using the optional features \emph{Voice ID} and \emph{Visual ID}, respectively. %
While Voice ID has already been available for Echo smart speakers, 
Visual ID was introduced with the Echo Show~15 
and enables face detection and recognition 
to display user-dependent content \citep{amazon2021sciencevisualid}. %
Meanwhile, Visual~ID has been rolled out to Echo Show~8 (Gen.\,2) 
and Echo Show~10 (Gen.\,3) devices as well 
\citep{amazon:what-is-visualid-on-echo-show}.

\paragraph{User Accounts.} %

An Amazon account is mandatory for setting up the Echo Show~15. %
For additional users in a multi-person household, 
so-called \emph{Alexa profiles}  
can be created for both adults and kids \citep{amazon:alexa-profiles}. %
Since there is no additional authentication, 
any individual with close spatial or immediate physical access 
to the Echo Show~15 may issue voice commands 
or perform actions via touch gestures under an arbitrary profile. %
Therefore, even though enabled Voice ID or Visual ID may associate 
interactions to a certain profile, 
attributions to individuals must be made with caution. %

\paragraph{Cloud \& Apps.} %

Generally, Echo and Echo Show devices strongly rely on the Alexa cloud, 
as voice commands are processed remotely by the \acl{AVS} (\acs{AVS}) 
\citep{chung2017alexaecosystem,krueger2020using,youn2021echoshow}. %
Likewise, the cloud connection is required to synchronize 
user data across devices, including large files like photos 
which are uploaded to Amazon's online storage. %
For display-less Echo smart speakers, the companion app \emph{``Amazon Alexa''}
is essential to set up a network connection via Wi-Fi, 
while smart displays like the Echo Show~15 
can be set up directly using the touchscreen. %
After the initial setup, the Alexa companion app can be further used to, inter alia, 
engage in text and video conversations with other users, 
control connected smart home devices, 
install additional Alexa \emph{Skills} 
(i.e., add-on apps to enhance the capabilities of Alexa), 
or manage device settings and user data 
stored in the vendor's cloud, 
such as calendars, contacts, to-do and shopping lists, 
or a history of past Alexa voice commands. %
Another companion app called \emph{``Amazon Photos''} 
is available for managing the photos in Amazon's online storage. %

\subsection{Related Work}
\label{background:related-work}

\citet{chung2017alexaecosystem} proposed a general approach for 
examining Alexa-powered smart devices forensically 
and suggested to organize the analysis based on the 
components involved in the Alexa ecosystem, i.e., 
Alexa-enabled devices, 
clients like companion apps or web browsers,
the Alexa cloud, 
and the network communication within this ecosystem. %
Their analysis resulted in an 
unofficial Alexa \acs{API} description as well as \emph{CIFT}, 
a proof-of-concept tool for cloud-based IoT forensics. %
\citet{youn2021echoshow} performed a chip-off on an Echo Show (Gen.\,2) device 
and extracted, inter alia, user credentials (i.e., Account ID, email address, and password) 
from the \acs{eMMC} to access remote artifacts in the vendor cloud. %
Furthermore, they refined the unofficial \ac{API} description 
of \citet{chung2017alexaecosystem} 
by functionalities that are specific for smart displays.
\citet{krueger2020using} investigated the Alexa \ac{API} and the
cloud artifacts of second- and third-generation Echo Dot devices, 
and found that certain information can be found in multiple
places which facilitates the recognition of manipulation attempts. %

Many previous work on Alexa forensics 
focused on detecting hardware debug interfaces
for obtaining file system access 
\citep{%
    clinton2016survey,
    hyde2017alexa,
    vasile2019breakingall,
    vanderpot2019echohackingwiki,
    pawlaszczyk2019alexa,
    riverloop2020echoeemc}. %
Identifying such interfaces on \aclp{PCB} usually requires manual probing
as available ports, pins, and \aclp{TP} are not necessarily annotated
and the board of each device has a different layout. %
In previous Echo models, different kinds of debug interfaces have been identified
which led to file system access (e.g., JTAG, UART, eMMC, ISP). %
For more recent devices of Amazon's Echo family, however, no debug interfaces were
discovered that allow reading the flash memory chip non-invasively, 
making a destructive chip-off procedure necessary. %

The vendor's cloud is of fundamental value for forensic investigations, 
as Echo devices store most of their user data remotely 
\citep{chung2017alexaecosystem,krueger2020using,youn2021echoshow}. %
Accessing this data, however, is not trivial, because nowadays credentials for
the Alexa \ac{API} are only stored as encrypted tokens on the device
and within companion apps \citep{hyde2017alexa, hutchinson:2022:wearables}. %
Although \citet{olufohunsi:2020:alexa} assumed that the encryption key is stored
in the same database as the encrypted tokens, a decryption has not been reported yet, 
making investigators depend on either the cooperation of the user or Amazon 
during data acquisition. %

\section{Hardware Examination}
\label{hardware-examination}

In July 2022 and February 2023, we purchased two Echo Show~15 devices 
that had the same \acl{MLB} (\acs{MLB}) build number (``30-006507 REV01''). %
Below, we present our hardware-related findings wrt. the smart display's 
Micro-USB port (\Cref{hardware-examination:usb}), 
\acp{PCB} (\Cref{hardware-examination:pcb}), 
UART port (\Cref{hardware-examination:pcb:uart}),
Micro-HDMI port (\Cref{hardware-examination:pcb:hdmi}), 
as well as a yet undocumented pinout for the eMMC interface 
which we revealed (\Cref{hardware-examination:pcb:emmc}). %

\subsection{Micro-USB Port}
\label{hardware-examination:usb}

When connecting the Echo Show 15 to a computer via its Micro-USB port 
and running \texttt{lsusb}, 
it is not recognized as a USB device. %
For Echo Dot smart speakers, 
a button combination has been identified in prior work 
\citep{micaksica2017exploringechodot2,vanderpot2019echohackingwiki} 
that boots the device into \emph{fastboot} mode and makes it
available as fastboot device. %
Fastboot is a protocol for communicating with the Android bootloader. %
For the Echo Show~15, we found a similar button combination:
Pressing the volume down and power/privacy button concurrently while powered off, 
the screen shows the Amazon logo on a black background after about four seconds. %
However, pressing both the volume up and down button 
and then the power/privacy button leads to a factory reset. %
Booted into fastboot mode, the Echo Show~15 is 
recognized by \texttt{lsusb}, and the tool 
\texttt{fastboot} 
can read the bootloader variables 
which reveal that the bootloader is locked, 
the bootloader's version (``01.01.220125.215459''), 
the device's serial number (16 alphanumeric characters) 
and internal product name (``hoya''), 
and an incomplete partition table 
(\textsf{boot}, \textsf{system}, \textsf{vendor}, \textsf{odm}, \textsf{data}; 
cf. \Cref{tbl:filesystem:partitions} in \Cref{filesystem-artifacts:comparison-new-old}). %

\subsection{Printed Circuit Boards}
\label{hardware-examination:pcb}

\begin{figure}[t!]
  \centering
  \includegraphics[width=\linewidth]{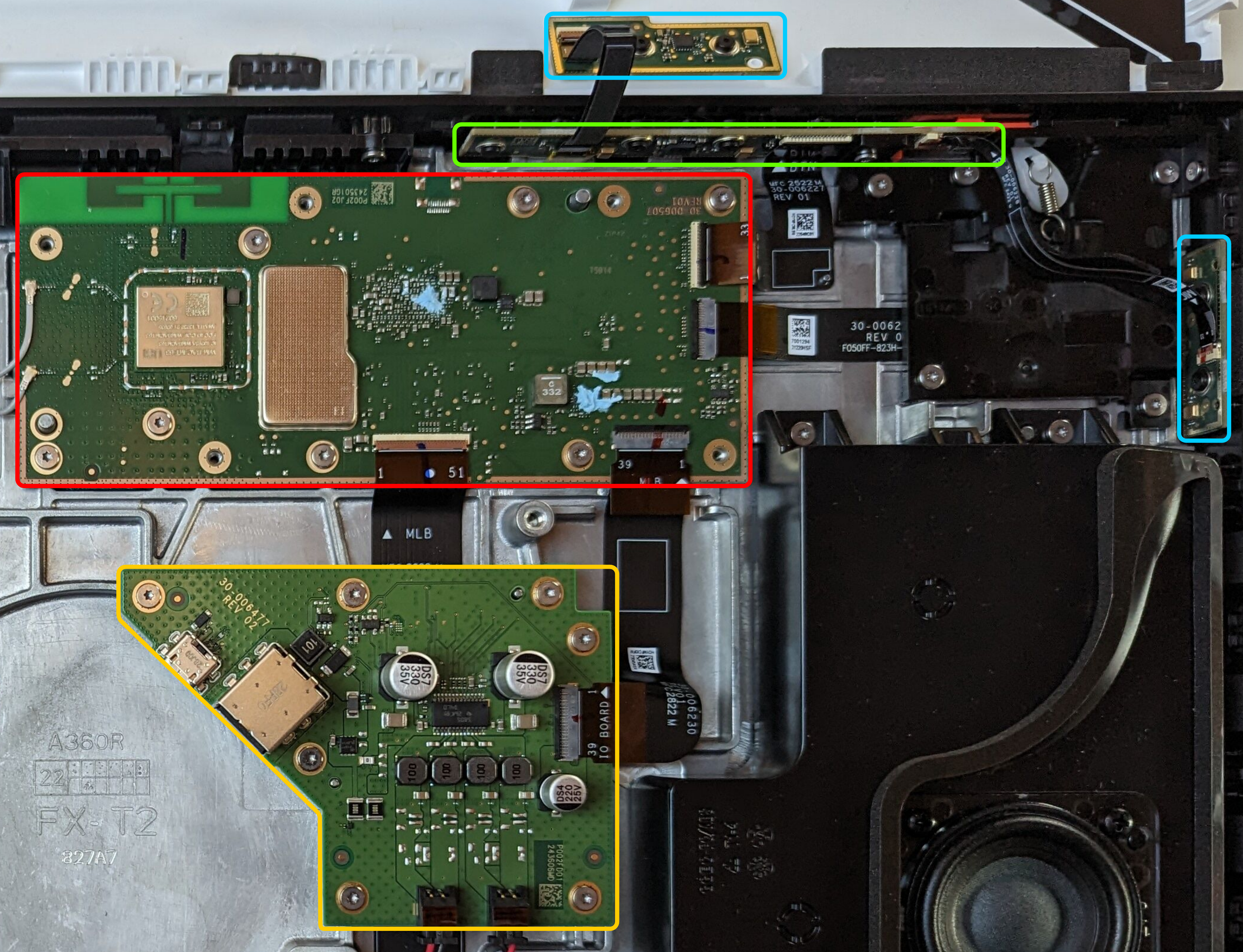}
  \caption{%
    Overview of \acp{PCB}: Main logic board (\protect\colorcirc{pcb_red}),
    power and audio (\protect\colorcirc{pcb_yellow}),
    buttons, camera cover slider and one microphone array
    (\protect\colorcirc{pcb_green}),
    and \acp{PCB} with a microphone array each (\protect\colorcirc{pcb_blue}).%
  }
  \label{fig:hardware:pcb:overview}
\end{figure}

Since fastboot via Micro-USB port did not allow further access, 
we unscrewed and removed the backplate of the Echo Show~15 
to take a look at the \acp{PCB} shown in \Cref{fig:hardware:pcb:overview}:
\begin{compactitem}
\item Two \acp{PCB}, each with a microphone array.
\item A \ac{PCB} 
    on the top side of the Echo Show 15
    which contains 
    a camera cover slider, 
    a power/privacy button, 
    two volume buttons (up/down), 
    a microphone array.
\item A \ac{PCB} 
    for power management and audio processing, 
    including a Micro-USB port.
\item The \ac{MLB} 
    shown in \Cref{fig:hardware:pcb:mlb} with 
    Amlogic PopcornA\allowbreak{}/POP1-C \acs{SoC} (cf. \Cref{background:echo-show-15}),
    2GB LPDDR4 RAM and 1GB LPDDR4x RAM from Samsung, 
    an \acs{eMMC} with 16GB (SD\-INBDG4-16G) 
    by SanDisk as BGA\,153 package,  
    a dual-band Wi-Fi (802.11a/b/g/n/ac) and
    Bluetooth 5.1 module by USI (WM-BAC-MT-63).
\end{compactitem}
For debugging and testing purposes, manufacturers often integrate
\aclp{TP} (\acsp{TP}) and ports 
for serial communication into their \acp{PCB}. %
In earlier Echo models, such interfaces were available
and allowed to access a shell, to dump the firmware by reading the flash
memory, or to boot from an external SD card \citep{clinton2016survey}. %
In later Echo models, however, 
those interfaces were either limited 
or eliminated, whereby a chip-off 
remained the last option to access file system data 
\citep{pawlaszczyk2019alexa, youn2021echoshow}. %

\subsubsection{UART Port}
\label{hardware-examination:pcb:uart}

\begin{figure}[t!]
  \centering
  \begin{subfigure}[b]{\linewidth}
    \centering
    \includegraphics[width=\linewidth]{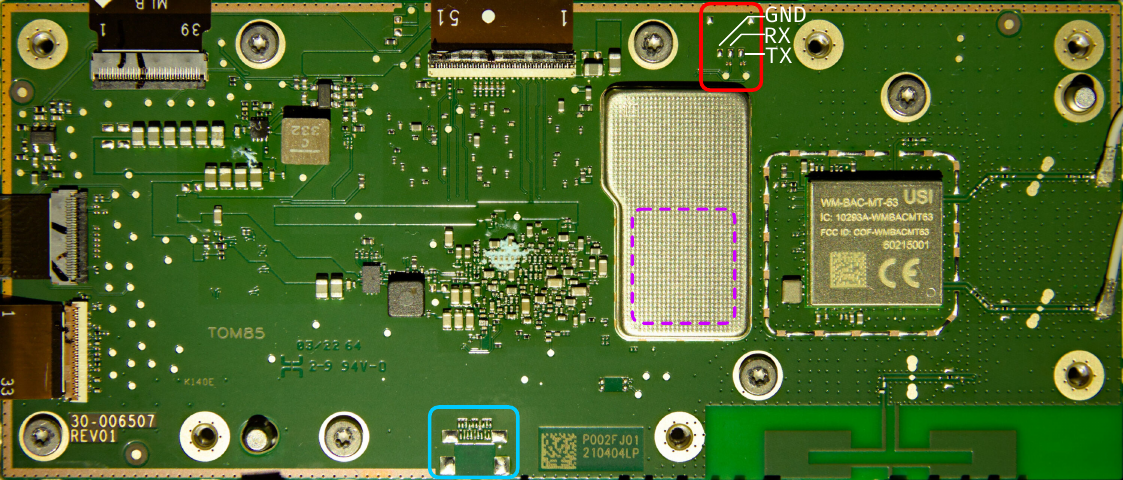}
    \caption{%
      Front: \acs{UART} (\colorcirc{pcb_red}), \acs{HDMI}
      (\colorcirc{pcb_blue}), \acs{eMMC}  
      (\colorcirc{pcb_purple}).
    }
    \label{fig:hardware:pcb:mlb:front}
  \end{subfigure}

  \vspace*{0.5em}

  \begin{subfigure}[b]{\linewidth}
    \centering
    \includegraphics[width=\linewidth]{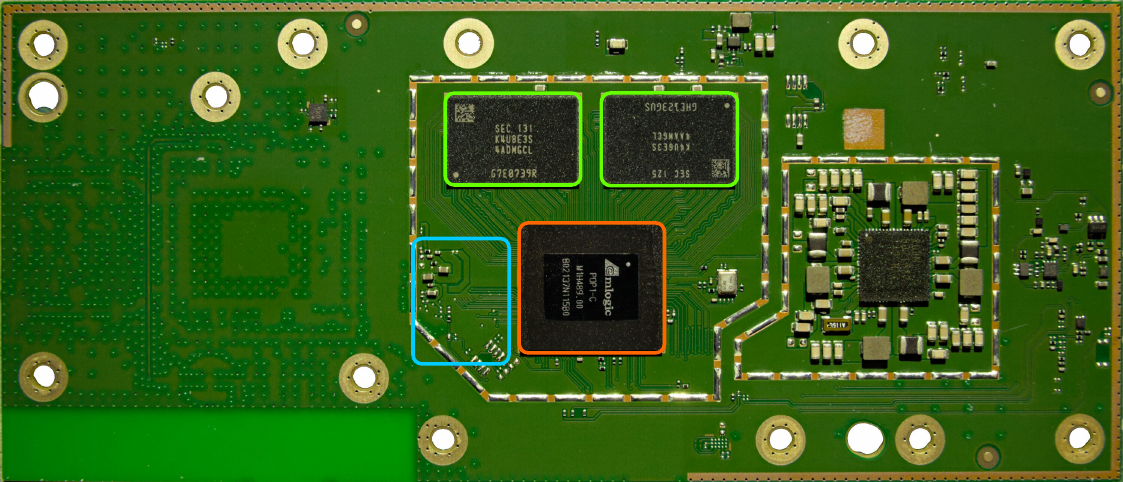}
    \caption{%
      Back: SoC (\colorcirc{pcb_orange}), RAM
      (\colorcirc{pcb_green}), area of \ac{eMMC} pinout
      (\colorcirc{pcb_blue}).
    }
    \label{fig:hardware:pcb:mlb:back}
  \end{subfigure}

  \caption{Main logic board (\acs{MLB}) of the Echo Show 15.}
  \label{fig:hardware:pcb:mlb}
\end{figure}

We compared the \ac{MLB} of our Echo Show~15  %
to internal photos of a pre-market version 
submitted to the \ac{FCC} \citep{fcc2021-2AXFL-4269} 
to identify missing components in the end product 
which may reveal debug ports. %
The front side of the pre-market 
\ac{MLB} contained a 3-pin connector, 
while our board only had the contacts the connector was soldered to 
(cf. \Cref{fig:hardware:pcb:mlb:front}). %
Considering the debug interfaces found in earlier Echo models, 
only \acs{UART} (\acl{UART}) uses three pins: 
Ground (GND), receive (RX), and transmit (TX). %
Using a multimeter, the contact without \ac{TP}
was identified as Ground, and both remaining contacts were measured around
3.3V with a little fluctuation (TX) and constant 3.1V (RX), respectively. %
For establishing a serial connection
from the Echo Show~15 to a computer via \ac{UART}, 
we set an UART-to-USB adapter to 3.3V 
and attached jump wires which were 
soldered to the identified contacts on the \ac{MLB}. %
We identified a baud rate of 921600 as fitting (\emph{8N1}: 
8 data bits, no parity bit and 1 stop bit), 
while earlier Echo devices had baud rates of 115200 or 912000 \citep{vasile2019breakingall}. %
Once the Echo Show~15 was powered on, 
the command-line tool \texttt{screen} printed the boot logs, 
including the logs of the bootloader \emph{U-Boot} 
(v2019.01, build ID ``jenkins-fireos\_main\_pie-patch-build-193097''), 
which was already used as bootloader in previous Echo devices
\citep{clinton2016survey, vasile2019breakingall}. %
The boot log was not followed by a login shell, 
as is sometimes the case for \ac{UART} interfaces. %
In addition to the information that the firmware image is signed 
and the number of device starts (\texttt{bootcount}), 
the boot log indicated to ``\textit{Hit Enter key to stop autoboot}'', 
which led to a bootloader shell. %
Although available commands were revealed by tab completion, 
all of them were blocked. %

\subsubsection{Micro-HDMI}
\label{hardware-examination:pcb:hdmi}

Another difference between the \ac{MLB} of the 
Echo Show 15 device filed to the \ac{FCC} and ours 
was a missing Micro-HDMI port, of which only the solder pads marked in
\Cref{fig:hardware:pcb:mlb:front} were left on the board of the end product. %
Our attempt to solder a \acs{HDMI} breakout board to the pads 
and connecting both a display and a video source 
yielded no observable response of the Echo Show~15. %
Therefore, the purpose of the Micro-HDMI port 
as well as its proper functioning in the end product 
remained unclear to us (cf. \Cref{discussion:limitations-future-work}).

\subsubsection{eMMC Interface}
\label{hardware-examination:pcb:emmc}

\begin{figure}[tb]
  \centering
  \includegraphics[width=0.9\linewidth]{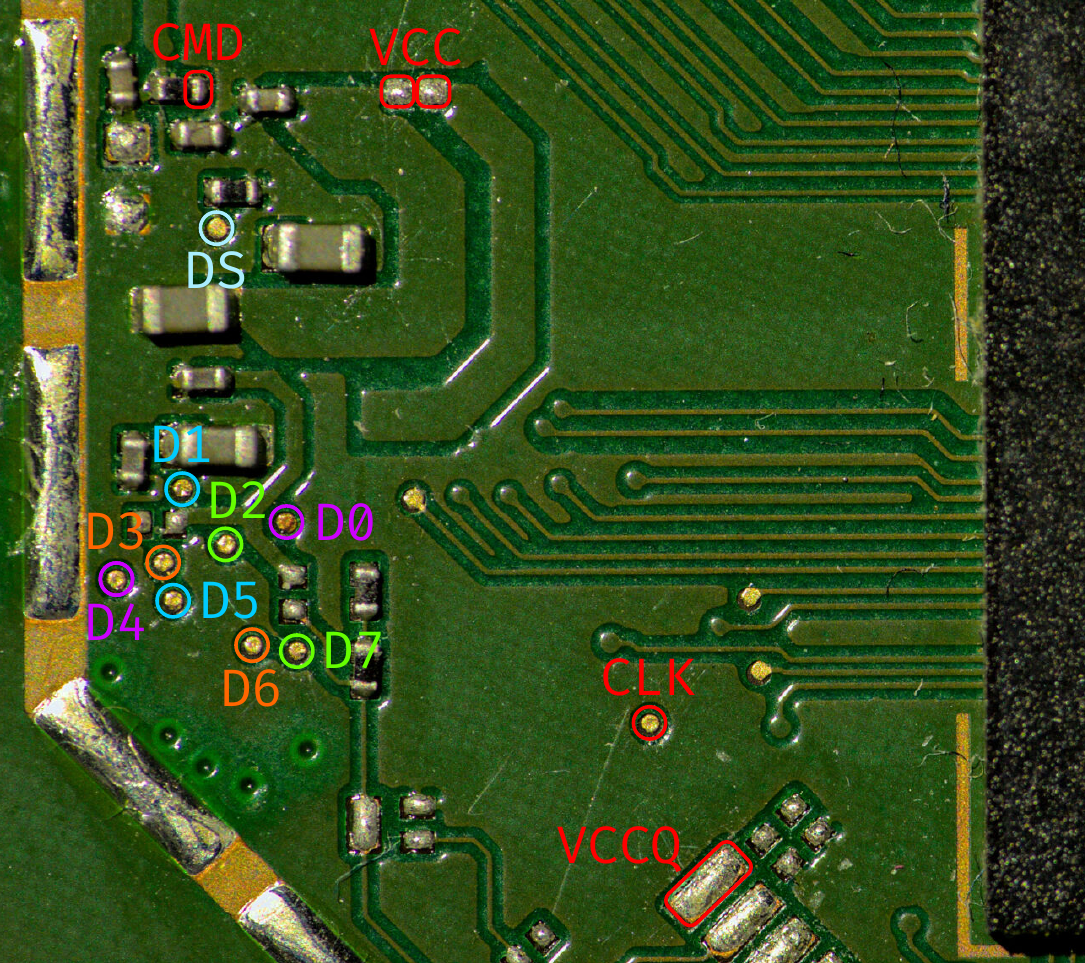}
  \caption{Unveiled pinout for the \ac{eMMC} interface.}
  \label{fig:hardware:pcb:mlb:emmc:pinout}
\end{figure}

\begin{figure}[t!b]
  \centering
  \includegraphics[width=0.65\linewidth]{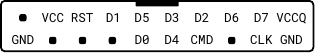}
  \caption{EasyJTAG pinout in \textit{EasyJTAG2/E-Socket} mode.}
  \label{fig:hardware:pcb:mlb:emmc:easyjtag-pinout}
\end{figure}

During our search for further debug interfaces, 
we used a logic analyzer to check the signals of \acp{TP} on the \ac{MLB}. %
Unlike the discovered asynchronous \ac{UART} interface, 
other interfaces need a clock for synchronous communication 
(e.g., \acs{JTAG}, \acs{ISP}, \acs{eMMC}). %
Although we found a clock signal, 
we could not identify protocol-specific \acp{TP} systematically. %
%
Since we had not found a non-invasive way to read out the \ac{eMMC} yet, 
we performed a chip-off. %
With the \ac{eMMC} contacts being exposed after removing the chip, 
we were able to measure continuity between them 
and a group of \acp{TP} on the back of the \ac{MLB} (cf. \Cref{fig:hardware:pcb:mlb:back}), 
including the clock mentioned above. %
\Cref{fig:hardware:pcb:mlb:emmc:pinout} shows that  
all relevant \ac{eMMC} channels (cf. \Cref{tbl:hardware:pcb:mlb:emmc:channels}) 
can be accessed through undocumented \acp{TP}. %
For Ground, one of the grounded mounting holes can be used.
\begin{table}[tb]
  \centering
  \rmfamily
  \scriptsize
  \caption{Relevant \ac{eMMC} channels for the Echo Show~15.}
  \label{tbl:hardware:pcb:mlb:emmc:channels}
  \begin{tabular}{@{}ll@{}}
  \toprule
  \textbf{Channel} & \textbf{Description} \\
  \midrule
  CLK & Clock signal \\
  CMD & Bidirectional command line \\
  D0\dots D7 & Bidirectional data lines \\
  VCC & Input voltage for flash storage (3.3V)  \\
  VCCQ & Input voltage for controller (1.8V) \\
  VSS(Q) & Ground \\
  \bottomrule
\end{tabular}
\end{table}
As our first Echo Show~15 no longer had an \ac{eMMC} chip, 
we continued our examination with the second device. %
To read its \ac{eMMC} chip, needle probes were placed on all identified \acp{TP} 
(cf. \Cref{tbl:hardware:pcb:mlb:emmc:channels} and \Cref{fig:hardware:pcb:mlb:emmc:pinout}). %
Although the \ac{eMMC} protocol can operate with 1, 4, or 8 data channels,
we only used the \texttt{D0}-channel in 1-bit mode due to space limitations. %
The probes were connected to an \emph{EasyJTAG Plus} box, 
a tool for reading various types of flash memory chips 
which is equipped with a 20-pin connector that is compatible with
a variety of sockets and extension cards. %
We connected the probes directly to the 20-pin connector %
according to the pin assignment provided by the \emph{EasyJTAG Classic Suite}
software (cf. \Cref{fig:hardware:pcb:mlb:emmc:easyjtag-pinout}). %
With the pinout mode ``EasyJTAG2/E-Socket'', 
a clock rate of 1~MHz,
an IO voltage of 1.8V, and a bus width of 1~bit,
all partitions could be dumped. %
Based on partial partition information found in the fastboot variables 
(cf. \Cref{hardware-examination:usb}), however, 
we recognized the absence of the \textsf{data} partition. %
As the remaining unmapped space was about the same size as the missing
partition, we dumped the unmapped space as well. %
Eventually, the ext4 \textsf{data} partition could be extracted 
from the unmapped space with \texttt{binwalk}. %

\section{Local Artifacts of the Echo Show 15}
\label{filesystem-artifacts}

Before reading the \acs{eMMC} chip of our second device non-invasively 
using the undocumented pinout for the \acs{eMMC} interface (cf. \Cref{hardware-examination:pcb:emmc}), 
we generated test data by using the smart display regularly. %
%
Below, we address how we generated the test data 
(\Cref{filesystem-artifacts:test-data-generation}), 
the layout of the partition table 
(\Cref{filesystem-artifacts:partition-table}), 
which artifacts known from previous Echo products 
(\Cref{filesystem-artifacts:comparison-new-old}) 
as well as such related to logs and interaction events 
(\Cref{filesystem-artifacts:logs-device-voice}) 
and Visual ID (\Cref{filesystem-artifacts:logs-visual})  
were found on the Echo Show~15,
and how credential tokens for the Alexa cloud can be decrypted 
(\Cref{filesystem-artifacts:token-database}). %

\subsection{Test Data Generation}
\label{filesystem-artifacts:test-data-generation}

We installed the Echo Show 15 in a separate room 
so that every stimulus to the camera and microphones could be documented. %
To correlate performed interactions and persisted artifacts, 
most interactions with the smart display were documented nearly to the second. %
The Echo Show~15 was used between February and August 2023 
with Fire OS versions 7.5.0.1 (PS7501/4131) -- 7.5.5.9 (PS7559/3534) 
as follows: %
Alexa was regularly asked for weather forecasts, news, 
and estimated travel times to work. %
Appointments have been created in the calendar, 
and a Google calendar was synchronized. %
The shopping and to-do lists were filled with items 
that were occasionally ticked off. %
The built-in browser Silk was used to browse the Web via touchscreen, 
and to store credentials for certain websites. %
Movies were watched using Amazon's Prime Video
as well as other services offered by Fire TV (e.g., YouTube). %
We created additional Alexa profiles and configured respective Voice IDs and Visual IDs. %
Since the Echo Show 15 responds to motion by waking up from a dimmed screen, 
we documented when a person was in the camera's view and whether their face was visible to the camera. %
The camera was used to take pictures and videos, 
as well as for video calls via the \emph{Amazon Alexa} companion app. %
Further, custom pictures were uploaded to the Echo Show~15 
via the \emph{Amazon Photos} companion app. %
Using the eponymous companion app, 
the capabilities of Alexa were extended by installing the Skill for Spotify. %
Finally, we linked a Wi-Fi smart camera to Alexa 
via the Skill ``Yi Home Camera'' 
for displaying the camera's live image on the Echo Show~15. %
A summary of artifacts generated by aforementioned activities is listed in
\Cref{tbl:cloud:action-findings}. %

\begin{table}[t!bh]
  \centering
  \rmfamily
  \scriptsize
  \caption[Local and remote artifacts of usage activities]{Local/remote artifacts of usage activities, as \emph{known} from other Echo devices and \emph{found} on the Echo Show~15 (*\,= volatile).}
  \label{tbl:cloud:action-findings}
  \begin{tabular}{@{}l@{\hspace{0.2\tabcolsep}}c@{\hspace{0.6\tabcolsep}}c@{\hspace{0.6\tabcolsep}}l@{}}
  \toprule
  \textbf{Artifacts} & \textbf{Known} & \textbf{Found} & \textbf{Location} \\
  \midrule
  User data & \fullcirc & \fullcirc & \ac{API}, Echo (partial) \tabularnewline
  Settings & \fullcirc & \fullcirc & \ac{API} \tabularnewline
  Wi-Fi credentials & \fullcirc & \fullcirc & Echo \tabularnewline
  Voice requests/responses & \fullcirc & \fullcirc & \ac{API}, Echo (cache) \tabularnewline
  Video calls & \fullcirc & \fullcirc & \ac{API} (metadata) \tabularnewline
  Browser (history, cred., cookies) & \fullcirc & \fullcirc & Echo \tabularnewline
  Lists (Shopping, To-Do) & \fullcirc & \fullcirc & \ac{API} \tabularnewline
  Calendars (Alexa, Google) & \fullcirc & \fullcirc & \ac{API} \tabularnewline
  Photos \& videos (w/ metadata) & \fullcirc & \fullcirc & \ac{API} \tabularnewline
  Connected devices & \fullcirc & \fullcirc & \ac{API} \tabularnewline
  Songs played on Spotify & \fullcirc & \fullcirc & \ac{API} \tabularnewline
  Conversations w/ Alexa users & \fullcirc & \fullcirc & \ac{API} \tabularnewline
  Local notes for local users & \emptycirc & \emptycirc & -- \tabularnewline
  Local users (Voice ID, Visual ID) & \emptycirc & \fullcirc & \ac{API} \tabularnewline
  Prime Video history & \emptycirc & \fullcirc & Echo \tabularnewline
  Visual ID & \emptycirc & \fullcirc & Echo (logs*) \tabularnewline
  Privacy mode; close camera & \emptycirc & \fullcirc & Echo (logs*) \tabularnewline
  Use of Fire TV apps (e.g., YouTube) & \emptycirc & \fullcirc & Echo (cache*) \tabularnewline
  \bottomrule
\end{tabular}
\end{table}

\subsection{Partition Table of Fire OS}
\label{filesystem-artifacts:partition-table}

Using the non-invasive method described in \Cref{hardware-examination:pcb:emmc}, 
a file system dump with the partitions in \Cref{tbl:filesystem:partitions} 
was obtained from the Echo Show~15. %
A partition table found in the log file
\textsf{cache:}\path{/recovery/last_kmsg} 
helped to label the partitions. %
User-related data is stored on the \textsf{data} partition.

\subsection{Known Artifacts from Previous Works}
\label{filesystem-artifacts:comparison-new-old}

Local artifacts of forensic relevance have been identified in prior work 
for several previous Echo products (cf. \Cref{background:related-work}). %
\citet{youn2021echoshow} compiled a comprehensive list of 
user-related data on the Echo Show (Gen.\,2) 
based on which we found, for example, 
Wi-Fi credentials, log files indicating voice interactions and camera usage, 
metadata of taken photos and videos, and browser data on the Echo Show~15. %
Most of the remaining artifacts (cf. \Cref{tbl:cloud:action-findings}), 
such as user-related data, voice commands, lists, calendars, 
device information, or photos and videos, 
however, are stored in the cloud and 
synchronized with the companion apps or other connected Echo devices 
\citep{chung2017alexaecosystem,krueger2020using}. %
An extensive and detailed overview of all our local and remote artifact findings 
for the Echo Show~15, including a comparison with related work, 
is given in \Cref{tbl:cloud:findings}. %
In the following subsections, we will therefore focus 
on device-specific artifacts discovered on the Echo Show 15. %

\begin{table}[t!bh]
  \centering
  \rmfamily
  \scriptsize
  \caption[%
    Partitions found on the \ac{eMMC}.%
  ]{%
    Partitions found on the \ac{eMMC} (*\,= listed in fastboot).
  }
  \label{tbl:filesystem:partitions}
  \begin{tabular}{@{}lrll@{}}
  \toprule
  \textbf{Partition} & \textbf{Size} & \textbf{Offset} & \textbf{Size (Bytes)} \\
  \midrule
  bootloader & 4 MB & \texttt{0x00000000} & \texttt{0x000400000} \tabularnewline
  reserved & 8 MB & \texttt{0x02400000} & \texttt{0x000800000} \tabularnewline
  nvcfg & 4 MB & \texttt{0x02d00000} & \texttt{0x000400000} \tabularnewline
  tee & 8 MB & \texttt{0x03200000} & \texttt{0x000800000} \tabularnewline
  boot* & 24 MB & \texttt{0x03b00000} & \texttt{0x001800000} \tabularnewline
  recovery & 24 MB & \texttt{0x05400000} & \texttt{0x001800000} \tabularnewline
  logo & 4 MB & \texttt{0x06d00000} & \texttt{0x000400000} \tabularnewline
  misc & 1 MB & \texttt{0x07200000} & \texttt{0x000100000} \tabularnewline
  cri\_data & 2 MB & \texttt{0x07400000} & \texttt{0x000200000} \tabularnewline
  vendor* & 300 MB & \texttt{0x07700000} & \texttt{0x012c00000} \tabularnewline
  odm* & 8 MB & \texttt{0x1a400000} & \texttt{0x000800000} \tabularnewline
  system* & 3 GB & \texttt{0x1ad00000} & \texttt{0x0c2000000} \tabularnewline
  product & 12 MB & \texttt{0xdce00000} & \texttt{0x000c00000} \tabularnewline
  cache & 512 MB & \texttt{0xddb00000} & \texttt{0x020000000} \tabularnewline
  data* & 10.7 GB & \texttt{0xfdc00000} & \texttt{0x2ad800000} \tabularnewline
  \bottomrule
\end{tabular}
\end{table}

\subsection{Logs and Interaction Events}
\label{filesystem-artifacts:logs-device-voice}

Fire OS uses Android's 
\emph{DropBoxManager}\footnote{\href{https://developer.android.com/reference/android/os/DropBoxManager.html}{developer.android.com/reference/android/os/DropBoxManager.html}} 
to write system logs on the \textsf{data} partition in
\path{/system/dropbox/} and \path{/logd/} stored as ZIP archives named
``\texttt{Log.}{\footnotesize\emph{\string{category\string}}}\allowbreak{}\texttt{@}{\footnotesize\emph{\string{unixtimestamp\string}}}%
\texttt{.txt.zip}'' using the categories 
\emph{crash}, \emph{events}, \emph{kernel}, \emph{main},
\emph{metrics}, \emph{system} and \emph{vitals}. %
Prior work showed that these logs allow to draw conclusions about
user interactions \citep{youn2021echoshow}. %
Calling the \emph{wake word} (i.e., voice command required to start interaction with Alexa)
is logged in ``\texttt{Log.system.*}'' as \texttt{WAKE\_WORD} event. %
Every button press is recorded as \texttt{BUTTON\_EVENT}, 
and every touch on the screen as \texttt{TOUCH\_EVENT}. %
On the Echo Show~15, Alexa can be prevented from listening and watching 
by pressing the power/privacy button, 
or by closing the camera slider 
(cf. \Cref{fig:background:echo-show-15:buttons}), 
resulting in 
\texttt{PRIVACY\_\allowbreak{}MODE\_\allowbreak{}\{ON,OFF\}} and
\texttt{CAMERA\_\allowbreak{}\{EN,DIS\}ABLED} events
being logged in ``\texttt{Log.system.*}'', respectively.

\subsection{Visual ID Artifacts}
\label{filesystem-artifacts:logs-visual}

Due to its face recognition feature, 
the Echo Show~15 constantly observes the room via its camera,  
even when users are not actively interacting with the device. %
Interestingly, Visual ID does not identify users 
when the device was started without Internet connection 
and stays offline. %
However, if the device is started while connected 
to the Internet and then taken offline, 
Visual ID works. %
When the camera observes any movement, 
a \texttt{MOTION} event is logged 
in ``\texttt{Log.main.*}'', 
indicating whether the motion originated 
from a person and whether that person is enrolled in Visual ID, 
as well as the reliability of the detection 
via a face quality score and the \texttt{personId} of the recognized face,
which is also reported to the Alexa cloud. %
The database ``\path{/data/com.amazon.alexa.identity/databases/recognition}'' 
stores the \texttt{personId} of users enrolled in Visual ID in the table
\emph{FaceEnrolledProfilesRecognition}. %
Note that values in the column \emph{lastRecognizedTimeMillis} 
do not indicate the last time a user was recognized, 
but---in our case---more likely the last time the device was started. %
The main user, whose Amazon account is logged
in, can be identified by searching for their \texttt{personId} in the
\emph{account\_\allowbreak{}data\_\allowbreak{}key} column 
of the table \emph{account\_\allowbreak{}data} in the database
\textsf{data}:``\path{/data/com.amazon.imp/databases/map_data_storage_v2.db}''
to link it to their \texttt{directedId} from column
\emph{account\_\allowbreak{}data\_\allowbreak{}directed\_\allowbreak{}id}. %
The corresponding username of the \texttt{directedId} 
can be obtained via the 
Alexa \ac{API} (cf. \Cref{cloud-app:cloud:api-artifacts:userids}).

\subsection{Token Database}
\label{filesystem-artifacts:token-database}

While \citet{youn2021echoshow} could 
retrieve user credentials from the Silk browser, 
this was no longer possible for the Echo Show~15. %
Although the user is automatically logged in into the Amazon website, 
an explicit login was required to access 
cloud \acp{API} (cf. \Cref{cloud-app:cloud}) 
and the Alexa webpage \path{alexa.amazon.com}; 
the latter allowed to view data stored in the cloud, 
but was taken offline during our tests. %
In earlier Echo models, the database ``\path{map_data_storage.db}'' 
contained access tokens for communicating with the Alexa \ac{API}
\citep{chung2017alexaecosystem}. %
Eventually, it was succeeded by ``\path{map_data_storage_v2.db}'' 
in \textsf{data}:``\path{/data/com.amazon.imp/databases/}'', which
stores all tokens in encrypted form and 
is also present in the Alexa app (cf. \Cref{cloud-app:companion:alexa}). %
\citet{olufohunsi:2020:alexa}
assumed that the Base64-encoded value for
\textit{key\_\allowbreak{}encryption\_\allowbreak{}secret} in the table
\textit{encryption\_\allowbreak{}data} could be the encryption key, 
but did not determine the type of encryption in use. %
To identify the most likely encryption scheme, 
we searched for open-source projects from Amazon 
and found\footnote{\label{ftn:aws-encryption}\href{https://github.com/aws/amazon-s3-encryption-client-java}{github.com/aws/amazon-s3-encryption-client-java}} 
that the values in ``\path{map_data_storage_v2.db}'' were 
encrypted using AES with \acf{CBC}, PKCS\#5 padding, 
and an initialization vector length of 16. %
Based on this information and the unprotected encryption key, 
the ``refresh token'' for the Alexa \ac{API} could be decrypted, 
resulting in unrestricted access to cloud artifacts 
(cf. \Cref{cloud-app:cloud}, \Cref{tbl:cloud:findings}). 

\section{Remote Artifacts of the Echo Show 15} 
\label{cloud-app}

As the network traffic of the Echo Show~15 is encrypted, 
and Fire OS does not allow to set a proxy or to inject custom certificates, 
we analyzed the remaining sources for acquiring remote artifacts: 
the two companion apps \emph{Amazon Alexa} and \emph{Amazon Photos} 
(\Cref{cloud-app:companion}), 
and the Amazon/Alexa cloud (\Cref{cloud-app:cloud}). %

\subsection{Storage of Companion Apps on Smartphone}
\label{cloud-app:companion}

We set up a rooted \ac{AVD} (9.0) to access the local data 
stored by the two companion apps for the Echo Show~15 
in the storage of an emulated smartphone via \texttt{adb}.

\subsubsection{Alexa App}
\label{cloud-app:companion:alexa}

The private application storage of the Amazon Alexa companion app 
(version 2022.21; 2.2.487227.0) 
is located at \path{/data/data/com.amazon.dee.app/} 
and contains, inter alia, the following artifacts: %

\begin{filepaths}
  \item[\path{shared_prefs/service.identity.xml}] %
    Lists the user's name, email address, user-specific IDs, and
    the temporary \ac{API} access token. %
    Since the access token expires after one hour, 
    it is unlikely that it will be still valid after retrieval. %
  \item[\path{databases/map_data_storage_v2.db}] %
    This database is equivalent to the one found on the Echo Show~15 
    (cf. \Cref{filesystem-artifacts:token-database}). %
    It stores, inter alia, the access and refresh token 
    used for authentication against the Alexa \ac{API} in encrypted form. %
  \item[\path{app_webview/Application Cache/Cache/}] %
    Cached data from in-app Webviews (cf. \citet{chung2017alexaecosystem} for format). %

  \item[\path{shared_prefs/mobilytics.session-storage.xml}] %
    Reveals the start and end timestamps of when the app was last used. %
\end{filepaths}

\subsubsection{Photos App}
\label{cloud-app:companion:photos}

The private application storage of the Amazon Photos companion app 
(version 2.1.0.107.0-aosp-902005930g) is located at
\path{/data/data/com.amazon.clouddrive.photos/} and contains, inter alia, 
the following artifacts: %

\begin{filepaths}
  \item[\path{databases/discovery_database_*}]
    Lists files that were uploaded in the app, incl.  
    file size, resolution, time of upload, time when the photo was taken, and MD5 hash. 
  \item[\path{databases/map_data_storage.db}]
    This database appears to be the predecessor of ``\path{map_data_storage_v2.db}''
    from the Alexa app, as the access token and refresh token for the Alexa \ac{API} 
    were still stored unencrypted during our tests.
  \item[\path{cache/image_manager_disk_cache/}]
    Directory containing cached images. While the pictures that were uploaded
    by the user kept their \acs{EXIF} metadata, the photos taken with the
    Echo Show~15's camera did not contain any information about when or with what device a
    picture was taken. Hidden pictures, as well as pictures which were deleted via
    the app, remain in the cache.
  \item[\path{databases/metadata_cache_database_*}] 
    The column \textit{data} of the table \textit{cache\_\allowbreak{}data} contains 
    \acs{JSON}-formatted metadata for cached images in
    ``\path{cache/image_manager_disk_cache/}'', such as filename, size, MD5 hash,
    capture time, and whether the image was put in the trash. The value of
    \texttt{createdBy} indicates whether a file was uploaded via the
    Photos app (\textit{Prime Photos Android}) or taken by the Echo Show
    (\textit{Knight Photos}).
    For uploaded photos taken with a smartphone or camera, 
    \acs{EXIF} metadata was retrievable. %
\end{filepaths}

\subsection{Amazon and Alexa Cloud}
\label{cloud-app:cloud}

Using an emulated smartphone (cf. \Cref{cloud-app:companion}), 
the network traffic of the companion apps has been intercepted 
by installing a custom root certificate in Android. %
The remote artifacts identified for the Echo Show~15, 
which are accessible through vendor \acp{API}, 
are summarized in \Cref{tbl:cloud:findings}, 
including a scope-wise comparison with related work 
for the individual local and remote artifacts.
Below, we first focus on the insights that we gathered about 
the different authorization methods (\Cref{cloud-app:cloud:authorization}) 
and kinds of user IDs (\Cref{cloud-app:cloud:api-artifacts:userids}) 
which are relevant for user identification when obtaining cloud artifacts. %
Afterward, we report on multimedia artifacts wrt.  
voice requests (\Cref{cloud-app:cloud:api-artifacts:voice-requests}) 
as well as photos and videos (\Cref{cloud-app:cloud:api-artifacts:photos}), 
before we address a GraphQL \ac{API} (\Cref{cloud-app:cloud:api-artifacts:graphql})   
that was not mentioned in the literature yet.

\subsubsection{Authentication}
\label{cloud-app:cloud:authorization}

The network analysis of the two companion apps 
showed that the vendor's \ac{API} endpoints are distributed 
across multiple hostnames and require distinct authentication methods 
(cf. \Cref{tbl:cloud:hostnames-auth}):
\begin{enumerate*}[label=(\roman*)]
  \item access token as Bearer in the \path{Authorization} header,
  \item access token in the \path{X-Amz-Access-Token} header, or
  \item session cookie.
\end{enumerate*}
Both the access token and the session cookies 
were also found in ``\path{map_data_storage_v2.db}'' 
(cf. \Cref{filesystem-artifacts:token-database}), 
but they are only valid for 1 and 24 hours, respectively.
Their renewal requires the refresh token 
found in ``\path{map_data_storage_v2.db}'' on the Echo Show~15
or within the companion apps'
local data (cf. \Cref{cloud-app:companion}).
The refresh token appears to only expire 
when the user logs out of an app or device. %

\begin{table}[tb]
  \centering
  \rmfamily
  \scriptsize
  \caption{Authorization methods depending on hostname (in our case).}
  \label{tbl:cloud:hostnames-auth}
  \begin{tabular}{@{}ll@{}}
\toprule
\textbf{Authorization Method}       & \textbf{Hostname} \\
\midrule
\texttt{Authorization} header       & \texttt{api.\{amazon,amazonalexa\}.com} \\
\arrayrulecolor{black!20}\cmidrule{1-2}\arrayrulecolor{black}
Session cookies                     & \texttt{\{alexa,skills-store,www\}.amazon.\{com,de,...\}}; \\
                                    & \path{alexa-comms-mobile-service.amazon.com} \\
\arrayrulecolor{black!20}\cmidrule{1-2}\arrayrulecolor{black}

\texttt{X-Amz-Access-Token} header  & \texttt{\{cdws.eu-west-1,drive\}.amazonaws.com} \\
\bottomrule
\end{tabular}

\end{table}

\subsubsection{User IDs}
\label{cloud-app:cloud:api-artifacts:userids}

Persons in the user's contact list and users with a local profile are assigned an ID. 
The various user IDs listed in \Cref{tbl:cloud:person-ids}, 
which are used by Amazon for different purposes, 
can be found not only in \ac{API} responses 
but also in files on the Echo Show~15 or within the
companion apps (cf. \Cref{tbl:cloud:findings}).
While \textit{customerId} and \textit{directedId} are stored together with
account-wide personal data, \textit{personId} is used to distinguish local
user profiles.
\textit{commsId} and \textit{contactId} are used in contact and conversational
contexts. %
The \ac{API} \path{/alexa-privacy/apd/rvh/persons-in-household} stores names of
local user profiles, which may have configured a Voice ID or Visual ID,
including their \textit{personId}.

\begin{table}[b!h]
  \centering
  \rmfamily
  \scriptsize
  \caption{Amazon's user-related IDs and their structure (in our case).}
  \label{tbl:cloud:person-ids}
  
\begin{threeparttable}
\begin{tabularx}{\linewidth}{@{}lX@{}}
  \toprule
  \textbf{Name} & \textbf{Structure} \\
  \midrule
  customerId & \textit{\{14-uppercase-alphanumerics\}} \tabularnewline
  directedId & amzn1.account.\textit{\{28-uppercase-alphanumerics\}} \tabularnewline
  commsId & amzn1.comms.id.person.amzn1$\sim$\textit{\{directedId\}} \tabularnewline
  contactId & \textit{\{uuid4\}} \tabularnewline
  personId\tnote{1} & amzn1.actor.person.did.\textit{\{72-uppercase-alphanumerics\}} \tabularnewline
  personIdV2 & amzn1.actor.person.oid.\textit{\{13-or-14-uppercase-alphanumerics\}} \tabularnewline
  \bottomrule
\end{tabularx}

\begin{tablenotes}
  \item[1] \textit{personId} is also often used as alias for
    \textit{directedId} or \textit{personIdV2}
\end{tablenotes}
\end{threeparttable}

\end{table}

\subsubsection{Voice Requests}
\label{cloud-app:cloud:api-artifacts:voice-requests}

All voice requests issued to the Echo Show~15 
are stored by default for indefinite time in the Alexa cloud.
Each record contains the device on which the request was received, a timestamp,
and a transcript what Alexa understood. 
Also, the intent which was derived from the user's request 
as well as IDs of resources that may get updated as a
consequence of the voice command are logged.
If Alexa was able to identify a person based on Voice ID, 
the associated \emph{personIdV2} is also saved (cf. \Cref{tbl:cloud:person-ids}).
Other user actions (e.g., via touch input), 
are not assigned to a user profile, 
even if the user was recognized by Visual~ID.
As Alexa repeatedly assigned requests of a person without enabled Voice ID to
another user's profile (with configured Voice ID) during our tests, although
those persons differed noticeably in their vocal pitch, 
any attribution by Alexa must be used with caution. 
If available, the recording of the request should be used for verification.
While these recordings can contain background noise or be deleted by the user, 
the associated metadata may still be available
\citep{krueger2020using}, incl. a transcript of the user's inquiry.

\subsubsection{Photos and Videos}
\label{cloud-app:cloud:api-artifacts:photos}

For a list of users' photos and videos, including metadata, 
the \ac{API} \path{cdws.eu-west-1.amazonaws.com/drive/v1/search} 
could be queried in our case. 
The individual unmodified files with preserved metadata 
could be fetched from
\path{content-eu.drive.amazonaws.com/v2/download/signed/}{\footnotesize\emph{\string{id\string}}}.

\subsubsection{GraphQL}
\label{cloud-app:cloud:api-artifacts:graphql}

The Alexa app queries a GraphQL API at \path{alexa.amazon.de/nexus/v1/graphql}. %
All observed requests were device-related and contained information
about our user's Alexa-enabled devices (Echo Show 15) 
and the Wi-Fi smart camera that we connected to Alexa 
(cf. \Cref{filesystem-artifacts:test-data-generation}). %
Further, we found indicators of GraphQL usage on the Echo Show~15, 
as the path ``\path{api/profile/graphql}'' and a GraphQL query were located
in the application \textsf{system}:\path{/system/priv-app/com.amazon.alexa.identity/com.amazon.alexa.identity.apk} 
using the command-line tool \texttt{strings}. %
We manually replicated all REST and GraphQL \ac{API} requests 
of the companion apps 
(cf. \Cref{tbl:cloud:findings}) 
using the tokens 
discovered on the Echo Show~15 (cf. \Cref{filesystem-artifacts:token-database}),   
but not GraphQL \ac{API} requests of the Echo Show~15 itself, 
as the ``csrf check failed'' 
(cf. \Cref{discussion:limitations-future-work}).

\section{Discussion}
\label{discussion}

In this section, we discuss 
the relevance of our findings to practitioners (\Cref{discussion:practical-relevance}), 
our artifacts coverage in light of related work (\Cref{discussion:related-work}), 
as well as limitations and future work (\Cref{discussion:limitations-future-work}).

\subsection{Practical Relevance of Findings}
\label{discussion:practical-relevance}

Since data acquisition has become increasingly challenging 
for even well-known devices like smartphones due to encryption, 
if not impossible without users' cooperation, 
acquiring data from unconventional devices during real-world investigations 
is unsurprisingly non-trivial.
In many cases, as outlined methodically by \citet{stachak2024nyon}, 
``\emph{manual acquisition}'' approaches via the user interface 
not only provide rather limited access to relevant artifacts, if any, 
but also jeopardize forensic soundness. %
While file system access via ``\emph{OS-based}'' interfaces 
is often restricted or not available at all, 
``\emph{hardware-based}'' acquisition techniques, if applicable, 
require specific knowledge and experience; 
in the case of chip-off, as a last resort, a certain willingness to take risks 
is required due to the possibility of complete data loss.  %

All aforementioned obstacles to examining unconventional devices 
in both single- and multi-source analysis contexts 
apply to the Echo Show~15 and its extended ecosystem consisting of companion apps and vendor cloud. %
Combining our individual findings, however, results in a comprehensive 
data acquisition strategy for Echo Show~15 devices, covering both local and remote artifacts 
(cf. \Cref{fig:discussion:methods-artifacts}): 
\begin{figure}[b!ht]
 \centering
 \includegraphics[width=\linewidth]{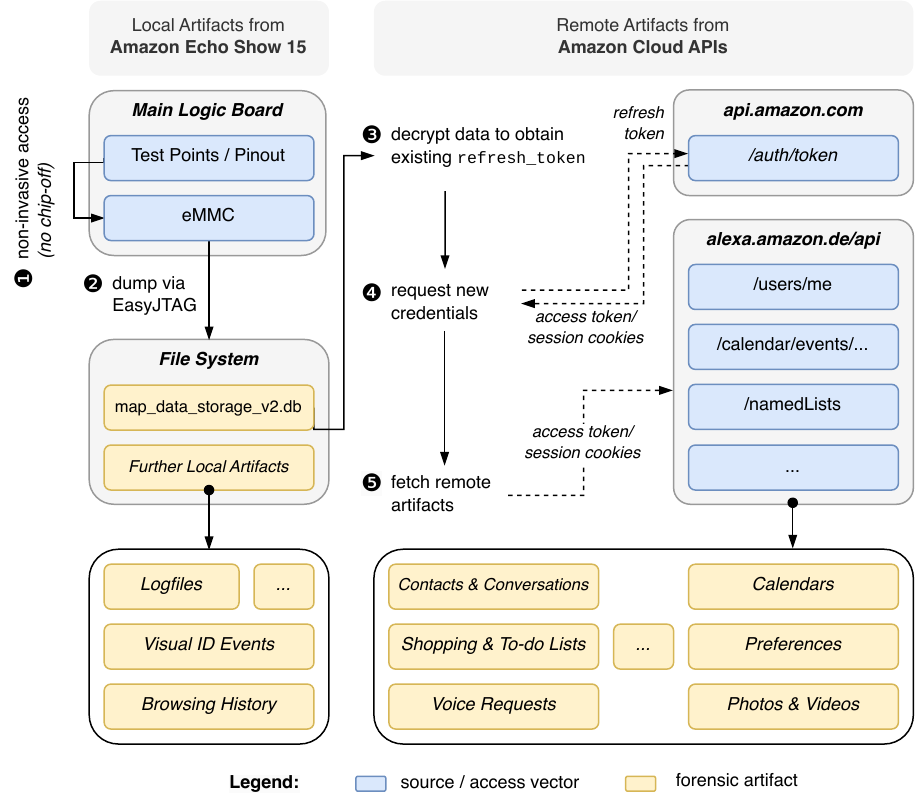}
 \caption{Data acquisition strategy for the Echo Show~15.}
 \label{fig:discussion:methods-artifacts}
\end{figure}
Based on the undocumented pinout for the eMMC interface 
that we identified (cf. \Cref{hardware-examination:pcb:emmc}), 
forensic investigators are now able to dump the file system non-invasively 
(\ballnumber{1}, \ballnumber{2}). %
Knowing how to decrypt the \texttt{refresh\_token} (\ballnumber{3}; cf. \Cref{filesystem-artifacts:token-database}), 
which seems to be indefinitely valid and is stored 
in the credential database ``\path{map_data_storage_v2.db}'' 
on the device together with the corresponding encryption key, 
forensic investigators can now request a temporary access token or session cookie (\ballnumber{4}) 
to authenticate themselves against the vendor \ac{API}, 
resulting in unrestricted access to user- and device-related remote artifacts 
stored in the vendor's cloud (\ballnumber{5}).

\subsection{Artifacts in Light of Related Work}
\label{discussion:related-work}

A detailed overview of all local 
and remote artifacts (cf. \Cref{filesystem-artifacts}, \Cref{cloud-app}) 
which we identified for the Echo Show~15 
are summarized in \Cref{tbl:cloud:findings}, 
including a comparison with related work 
regarding the coverage of individual artifacts. 
Additionally, we indicate which user IDs (cf. \Cref{tbl:cloud:person-ids})
are obtainable from certain artifacts, 
as these IDs are relevant for querying certain \acp{API} 
and attributing found artifacts to user accounts.
Except for two seemingly deprecated APIs for obtaining 
Wi-Fi credentials and user interactions with Alexa, 
as well as the \path{RKStorage1} artifact, 
we could either confirm the existence of known artifacts 
from previous work and earlier Echo devices 
\citep{chung2017alexaecosystem, hyde2017alexa, krueger2020using, olufohunsi:2020:alexa, youn2021echoshow}, 
or report on yet unaddressed artifacts across the multi-source environment of our examination 
consisting of the Echo Show~15, the two companion apps, 
and the Amazon cloud.

\begin{table*}[h!tb]
  \centering
  \notsotiny\rmfamily
  \caption[Summary of local and remote artifacts of the Echo Show~15 during our examination]{Summary of local and remote artifacts of the Echo Show~15 during our examination; incl. comparison with related work.}
  \label{tbl:cloud:findings}
  {
\setlength{\tabcolsep}{2pt}
\begin{threeparttable}
  \begin{tabular}{@{}llp{9.5cm}c@{\hspace{1.0\tabcolsep}}c@{\hspace{1.0\tabcolsep}}c@{\hspace{1.0\tabcolsep}}c@{\hspace{1.0\tabcolsep}}c@{\hspace{1.0\tabcolsep}}c@{}}
  \toprule
  \textbf{Source} & \textbf{Artifact(s)} & \textbf{Location (directory/file path, or API path)} & \textbf{[1]} & \textbf{[2]} & \textbf{[3]} & \textbf{[4]} & \textbf{[5]} & \textbf{We} \\
  \midrule
  Echo & Wi-Fi credentials & \path{/misc/wifi/WifiConfigStore.xml} & \emptycirc[0.8ex] & \halfcirc[0.8ex] & \emptycirc[0.8ex] & \emptycirc[0.8ex] & \emptycirc[0.8ex] & \fullcirc[0.8ex] \tabularnewline
  Echo & Cached images of searches & \path{/data/com.amazon.aria/cache/image_manager_disk_cache/} & \emptycirc[0.8ex] & \emptycirc[0.8ex] & \emptycirc[0.8ex] & \emptycirc[0.8ex] & \emptycirc[0.8ex] & \fullcirc[0.8ex] \tabularnewline
  Echo & Random screenshots & \path{/system_ce/0/snapshots/} & \emptycirc[0.8ex] & \emptycirc[0.8ex] & \emptycirc[0.8ex] & \emptycirc[0.8ex] & \emptycirc[0.8ex] & \fullcirc[0.8ex] \tabularnewline
  Echo & Random browser screenshots & \path{/data/com.amazon.cloud9/app_textures/} & \emptycirc[0.8ex] & \emptycirc[0.8ex] & \emptycirc[0.8ex] & \emptycirc[0.8ex] & \emptycirc[0.8ex] & \fullcirc[0.8ex] \tabularnewline
  Echo & Prime Video Watch History & \path{/data/com.amazon.avod/files/databases/dbplaybackhistory} & \emptycirc[0.8ex] & \emptycirc[0.8ex] & \emptycirc[0.8ex] & \emptycirc[0.8ex] & \emptycirc[0.8ex] & \fullcirc[0.8ex] \tabularnewline
  Echo & Last interaction by voice & \path{/data/amazon.speech.sim/shared_prefs/user_activity_prefs.xml} & \emptycirc[0.8ex] & \emptycirc[0.8ex] & \emptycirc[0.8ex] & \emptycirc[0.8ex] & \emptycirc[0.8ex] & \fullcirc[0.8ex] \tabularnewline
  Echo & Known devices & \multicolumn{1}{p{8cm}}{\path{/data/com.amazon.alexahybridremoteskill/files/customerHomeRegistry.db};\newline \path{/data/com.amazon.gloria.smarthome/shared_prefs/SmartHomeEntityCache.xml}} & \emptycirc[0.8ex] & \emptycirc[0.8ex] & \emptycirc[0.8ex] & \emptycirc[0.8ex] & \emptycirc[0.8ex] & \fullcirc[0.8ex] \tabularnewline
  Echo & Picture of connected Wi-Fi camera & \path{/data/com.amazon.cardinal/cache/} & \emptycirc[0.8ex] & \emptycirc[0.8ex] & \emptycirc[0.8ex] & \emptycirc[0.8ex] & \emptycirc[0.8ex] & \fullcirc[0.8ex] \tabularnewline
  Echo & Encrypted API credentials & \path{/data/com.amazon.imp/databases/map_data_storage_v2.db}\colorcirc{pcb_red}\colorcirc{pcb_green}\colorcirc{pcb_blue} & \halfcirc[0.8ex] & \emptycirc[0.8ex] & \emptycirc[0.8ex] & \fullcirc[0.8ex] & \fullcirc[0.8ex] & \fullcirc[0.8ex] \tabularnewline
  Echo & User data \& internal IDs & \path{/securedStorageLocation/com.amazon.alta.h2clientservice/databases/alta.h2clientservice.db} & \emptycirc[0.8ex] & \emptycirc[0.8ex] & \emptycirc[0.8ex] & \emptycirc[0.8ex] & \emptycirc[0.8ex] & \fullcirc[0.8ex] \tabularnewline
  Echo & Log files & \path{/system/dropbox/}\colorcirc{pcb_green} & \emptycirc[0.8ex] & \fullcirc[0.8ex] & \emptycirc[0.8ex] & \emptycirc[0.8ex] & \emptycirc[0.8ex] & \fullcirc[0.8ex] \tabularnewline
  Echo & Log files & \path{/logd/}\colorcirc{pcb_green} & \emptycirc[0.8ex] & \emptycirc[0.8ex] & \emptycirc[0.8ex] & \emptycirc[0.8ex] & \emptycirc[0.8ex] & \fullcirc[0.8ex] \tabularnewline
  Echo & Browser history, cookies, login data & \path{/data/com.amazon.cloud9/app_amazon_webview/amazon_webview/*} & \emptycirc[0.8ex] & \fullcirc[0.8ex] & \emptycirc[0.8ex] & \emptycirc[0.8ex] & \emptycirc[0.8ex] & \fullcirc[0.8ex] \tabularnewline
  Echo & Timestamp of last picture & \path{/data/com.amazon.zordon/shared_prefs/photobooth.xml} & \emptycirc[0.8ex] & \emptycirc[0.8ex] & \emptycirc[0.8ex] & \emptycirc[0.8ex] & \emptycirc[0.8ex] & \fullcirc[0.8ex] \tabularnewline
  Echo & Photo \& video metadata & \path{/data/com.amazon.zordon/databases/}\textrm{\tiny \textit{\{directedId\}}}\path{.mixtape.db} & \emptycirc[0.8ex] & \fullcirc[0.8ex] & \emptycirc[0.8ex] & \emptycirc[0.8ex] & \emptycirc[0.8ex] & \fullcirc[0.8ex] \tabularnewline
  Echo & Encrypted Visual ID photos & \path{/data/com.amazon.edgecvs/files/album/*/} & \emptycirc[0.8ex] & \emptycirc[0.8ex] & \emptycirc[0.8ex] & \emptycirc[0.8ex] & \emptycirc[0.8ex] & \fullcirc[0.8ex] \tabularnewline
  Echo & Event log & \path{/system/notification_log.db} & \emptycirc[0.8ex] & \emptycirc[0.8ex] & \emptycirc[0.8ex] & \emptycirc[0.8ex] & \emptycirc[0.8ex] & \fullcirc[0.8ex] \tabularnewline
  Echo & Timestamp of last boot & \path{/data/com.amazon.knight.calendar/shared_prefs/com.amazon.knight.calendar_preferences.xml} & \emptycirc[0.8ex] & \emptycirc[0.8ex] & \emptycirc[0.8ex] & \emptycirc[0.8ex] & \emptycirc[0.8ex] & \fullcirc[0.8ex] \tabularnewline
  Echo & Local user IDs with Visual ID & \path{/data/com.amazon.alexa.identity/databases/recognition}\colorcirc{pcb_green} & \emptycirc[0.8ex] & \emptycirc[0.8ex] & \emptycirc[0.8ex] & \emptycirc[0.8ex] & \emptycirc[0.8ex] & \fullcirc[0.8ex] \tabularnewline
  \arrayrulecolor{black!20}\cmidrule{1-9}\arrayrulecolor{black}
  Alexa app & User data \& internal IDs & \textrm{\tiny \textit{\{AApp\}}}\path{/shared_prefs/service.identity.xml}\colorcirc{pcb_purple}\colorcirc{pcb_red}\colorcirc{pcb_orange}\colorcirc{pcb_green} & \emptycirc[0.8ex] & \fullcirc[0.8ex] & \emptycirc[0.8ex] & \fullcirc[0.8ex] & \emptycirc[0.8ex] & \fullcirc[0.8ex] \tabularnewline
  Alexa app & Timestamp last app start & \textrm{\tiny \textit{\{AApp\}}}\path{/shared_prefs/SHARED_PREFS.xml}\colorcirc{pcb_red}\colorcirc{pcb_orange} & \emptycirc[0.8ex] & \fullcirc[0.8ex] & \emptycirc[0.8ex] & \fullcirc[0.8ex] & \emptycirc[0.8ex] & \fullcirc[0.8ex] \tabularnewline
  Alexa app & User data \& internal IDs & \textrm{\tiny \textit{\{AApp\}}}\path{/shared_prefs/SHARED_PREFS_IDENTITY.xml}\colorcirc{pcb_red}\colorcirc{pcb_orange} & \emptycirc[0.8ex] & \emptycirc[0.8ex] & \emptycirc[0.8ex] & \fullcirc[0.8ex] & \emptycirc[0.8ex] & \fullcirc[0.8ex] \tabularnewline
  Alexa app & Timestamps of last session & \textrm{\tiny \textit{\{AApp\}}}\path{/shared_prefs/mobilytics.session-storage.xml} & \emptycirc[0.8ex] & \emptycirc[0.8ex] & \emptycirc[0.8ex] & \emptycirc[0.8ex] & \emptycirc[0.8ex] & \fullcirc[0.8ex] \tabularnewline
  Alexa app & Session cookies & \textrm{\tiny \textit{\{AApp\}}}\path{/app_webview/Cookies} & \emptycirc[0.8ex] & \emptycirc[0.8ex] & \emptycirc[0.8ex] & \fullcirc[0.8ex] & \emptycirc[0.8ex] & \fullcirc[0.8ex] \tabularnewline
  Alexa app & Cached files of webviews & \textrm{\tiny \textit{\{AApp\}}}\path{/app_webview/Application Cache/Cache/} & \halfcirc[0.8ex] & \emptycirc[0.8ex] & \emptycirc[0.8ex] & \emptycirc[0.8ex] & \emptycirc[0.8ex] & \fullcirc[0.8ex] \tabularnewline
  Alexa app & Cached files of webviews & \textrm{\tiny \textit{\{AApp\}}}\path{/cache/org.chromium.android_webview/} & \emptycirc[0.8ex] & \emptycirc[0.8ex] & \emptycirc[0.8ex] & \emptycirc[0.8ex] & \emptycirc[0.8ex] & \fullcirc[0.8ex] \tabularnewline
  Alexa app & Encrypted API credentials & \textrm{\tiny \textit{\{AApp\}}}\path{/databases/map_data_storage_v2.db}\colorcirc{pcb_red}\colorcirc{pcb_green}\colorcirc{pcb_blue} & \halfcirc[0.8ex] & \emptycirc[0.8ex] & \emptycirc[0.8ex] & \fullcirc[0.8ex] & \fullcirc[0.8ex] & \fullcirc[0.8ex] \tabularnewline
  Alexa app & Shopping \& To-do lists & \textrm{\tiny \textit{\{AApp\}}}\path{/databases/DataStore.db}\colorcirc{pcb_purple} & \fullcirc[0.8ex] & \fullcirc[0.8ex] & \emptycirc[0.8ex] & \emptycirc[0.8ex] & \emptycirc[0.8ex] & \fullcirc[0.8ex] \tabularnewline
  Alexa app & User data \& internal IDs & \textrm{\tiny \textit{\{AApp\}}}\path{/databases/comms-core-identity-database}\colorcirc{pcb_red}\colorcirc{pcb_orange}\colorcirc{pcb_blue} & \emptycirc[0.8ex] & \emptycirc[0.8ex] & \emptycirc[0.8ex] & \emptycirc[0.8ex] & \emptycirc[0.8ex] & \fullcirc[0.8ex] \tabularnewline
  Alexa app & Conversations (encrypted?) & \textrm{\tiny \textit{\{AApp\}}}\path{/databases/comms.db} & \emptycirc[0.8ex] & \emptycirc[0.8ex] & \emptycirc[0.8ex] & \fullcirc[0.8ex] & \fullcirc[0.8ex] & \fullcirc[0.8ex] \tabularnewline
  Alexa app & Alarms, timers, transcribed requests & \textrm{\tiny \textit{\{AApp\}}}\path{/databases/RKStorage1} & \emptycirc[0.8ex] & \fullcirc[0.8ex] & \emptycirc[0.8ex] & \emptycirc[0.8ex] & \emptycirc[0.8ex] & \emptycirc[0.8ex] \tabularnewline
  \arrayrulecolor{black!20}\cmidrule{1-9}\arrayrulecolor{black}
  Photos app & Cached pictures & \textrm{\tiny \textit{\{PApp\}}}\path{/cache/image_manager_disk_cache/} & \emptycirc[0.8ex] & \emptycirc[0.8ex] & \emptycirc[0.8ex] & \emptycirc[0.8ex] & \emptycirc[0.8ex] & \fullcirc[0.8ex] \tabularnewline
  Photos app & API credentials (unencrypted) & \textrm{\tiny \textit{\{PApp\}}}\path{/databases/map_data_storage.db}\colorcirc{pcb_red} & \fullcirc[0.8ex] & \emptycirc[0.8ex] & \emptycirc[0.8ex] & \emptycirc[0.8ex] & \fullcirc[0.8ex] & \fullcirc[0.8ex] \tabularnewline
  Photos app & Metadata of uploaded pictures & \textrm{\tiny \textit{\{PApp\}}}\path{/databases/discovery_database_*} & \emptycirc[0.8ex] & \emptycirc[0.8ex] & \emptycirc[0.8ex] & \emptycirc[0.8ex] & \emptycirc[0.8ex] & \fullcirc[0.8ex] \tabularnewline
  Photos app & Metadata (also EXIF) of pictures & \textrm{\tiny \textit{\{PApp\}}}\path{/databases/metadata_cache_database_*}\colorcirc{pcb_purple} & \emptycirc[0.8ex] & \emptycirc[0.8ex] & \emptycirc[0.8ex] & \emptycirc[0.8ex] & \emptycirc[0.8ex] & \fullcirc[0.8ex] \tabularnewline
  \arrayrulecolor{black!20}\cmidrule{1-9}\arrayrulecolor{black}
  API & User data & \path{api.amazon.com/user/profile}\colorcirc{pcb_red} & \emptycirc[0.8ex] & \emptycirc[0.8ex] & \emptycirc[0.8ex] & \emptycirc[0.8ex] & \emptycirc[0.8ex] & \fullcirc[0.8ex] \tabularnewline
  API & User data & \path{alexa.amazon.de/api/users/me}\colorcirc{pcb_purple} & \emptycirc[0.8ex] & \emptycirc[0.8ex] & \emptycirc[0.8ex] & \emptycirc[0.8ex] & \emptycirc[0.8ex] & \fullcirc[0.8ex] \tabularnewline
  API & Wi-Fi credentials & \path{alexa.amazon.de/api/wifi/configs} & \fullcirc[0.8ex] & \emptycirc[0.8ex] & \emptycirc[0.8ex] & \fullcirc[0.8ex] & \fullcirc[0.8ex] & \textbf{--} \tabularnewline
  API & User interactions with Alexa & \path{alexa.amazon.de/api/activities} & \fullcirc[0.8ex] & \emptycirc[0.8ex] & \fullcirc[0.8ex] & \emptycirc[0.8ex] & \fullcirc[0.8ex] & \textbf{--} \tabularnewline
  API & Alexa app landing screen content & \path{alexa.amazon.de/api/content?personIdV2=}\textrm{\tiny \textit{\{personIdV2:did\}}}\colorcirc{pcb_purple}\colorcirc{pcb_blue} & \emptycirc[0.8ex] & \emptycirc[0.8ex] & \emptycirc[0.8ex] & \emptycirc[0.8ex] & \emptycirc[0.8ex] & \fullcirc[0.8ex] \tabularnewline
  API & Configured wake word & \path{alexa.amazon.de/api/wake-word} & \fullcirc[0.8ex] & \emptycirc[0.8ex] & \fullcirc[0.8ex] & \emptycirc[0.8ex] & \emptycirc[0.8ex] & \fullcirc[0.8ex] \tabularnewline
  API & Address, device information & \path{alexa.amazon.de/api/device-preferences}\colorcirc{pcb_purple} & \fullcirc[0.8ex] & \emptycirc[0.8ex] & \fullcirc[0.8ex] & \emptycirc[0.8ex] & \emptycirc[0.8ex] & \fullcirc[0.8ex] \tabularnewline
  API & Device list, capabilities & \path{alexa.amazon.de/api/devices-v2/device}\colorcirc{pcb_purple} & \halfcirc[0.8ex] & \emptycirc[0.8ex] & \fullcirc[0.8ex] & \halfcirc[0.8ex] & \halfcirc[0.8ex] & \fullcirc[0.8ex] \tabularnewline
  API & Device information & \path{alexa.amazon.de/api/bluetooth} & \fullcirc[0.8ex] & \emptycirc[0.8ex] & \fullcirc[0.8ex] & \emptycirc[0.8ex] & \emptycirc[0.8ex] & \fullcirc[0.8ex] \tabularnewline
  API & Shopping \& To-do lists & \path{pitangui.amazon.com/api/todos?type=}\textrm{\tiny \textit{\{TASK,SHOPPING\_ITEM\}}} & \fullcirc[0.8ex] & \emptycirc[0.8ex] & \emptycirc[0.8ex] & \emptycirc[0.8ex] & \emptycirc[0.8ex] & \textbf{--} \tabularnewline
  API & Shopping \& To-do list properties & \path{alexa.amazon.de/api/namedLists}\colorcirc{pcb_purple} & \emptycirc[0.8ex] & \fullcirc[0.8ex] & \fullcirc[0.8ex] & \emptycirc[0.8ex] & \emptycirc[0.8ex] & \fullcirc[0.8ex] \tabularnewline
  API & Shopping \& To-do list items & \path{alexa.amazon.de/api/namedLists/}\textrm{\tiny \textit{\{listId\}}}\colorcirc{pcb_purple} & \emptycirc[0.8ex] & \fullcirc[0.8ex] & \fullcirc[0.8ex] & \emptycirc[0.8ex] & \emptycirc[0.8ex] & \fullcirc[0.8ex] \tabularnewline
  API & Liveview enabled & \path{alexa.amazon.de/api/v1/devices/}\textrm{\tiny \textit{\{deviceAccountId\}}}\path{/settings/liveView} & \emptycirc[0.8ex] & \emptycirc[0.8ex] & \emptycirc[0.8ex] & \emptycirc[0.8ex] & \emptycirc[0.8ex] & \fullcirc[0.8ex] \tabularnewline
  API & Device information, online state & \path{cdws.eu-west-1.amazonaws.com/drive/v2/device-personalization/devices} & \emptycirc[0.8ex] & \emptycirc[0.8ex] & \emptycirc[0.8ex] & \emptycirc[0.8ex] & \emptycirc[0.8ex] & \fullcirc[0.8ex] \tabularnewline
  API & MAC address & \path{alexa.amazon.de/api/device-wifi-details} & \emptycirc[0.8ex] & \emptycirc[0.8ex] & \fullcirc[0.8ex] & \emptycirc[0.8ex] & \emptycirc[0.8ex] & \fullcirc[0.8ex] \tabularnewline
  API & Persons in household & \path{www.amazon.de/alexa-privacy/apd/rvh/persons-in-household}\colorcirc{pcb_blue} & \emptycirc[0.8ex] & \emptycirc[0.8ex] & \emptycirc[0.8ex] & \emptycirc[0.8ex] & \emptycirc[0.8ex] & \fullcirc[0.8ex] \tabularnewline
  API & User data & \path{alexa.amazon.de/api/household}\colorcirc{pcb_purple} & \fullcirc[0.8ex] & \emptycirc[0.8ex] & \fullcirc[0.8ex] & \emptycirc[0.8ex] & \emptycirc[0.8ex] & \fullcirc[0.8ex] \tabularnewline
  API & Alexa enabled devices & \path{alexa.amazon.de/api/phoenix} & \fullcirc[0.8ex] & \emptycirc[0.8ex] & \fullcirc[0.8ex] & \emptycirc[0.8ex] & \emptycirc[0.8ex] & \fullcirc[0.8ex] \tabularnewline
  API & Details of voice requests & \path{alexa.amazon.de/api/home}\colorcirc{pcb_purple} & \emptycirc[0.8ex] & \emptycirc[0.8ex] & \emptycirc[0.8ex] & \emptycirc[0.8ex] & \emptycirc[0.8ex] & \fullcirc[0.8ex] \tabularnewline
  API & Details of voice requests & \path{www.amazon.de/alexa-privacy/apd/rvh/customer-history-records?startTime=}\textrm{\tiny \emph{\{ts\}}}\path{&endTime=}\textrm{\tiny \emph{\{ts\}}}\colorcirc{pcb_purple}\colorcirc{pcb_blue} & \halfcirc[0.8ex] & \emptycirc[0.8ex] & \emptycirc[0.8ex] & \emptycirc[0.8ex] & \emptycirc[0.8ex] & \fullcirc[0.8ex] \tabularnewline
  API & Audio files of voice requests & \path{www.amazon.de/alexa-privacy/apd/rvh/audio?uid=}\textrm{\tiny \textit{\{utteranceId\}}} & \halfcirc[0.8ex] & \emptycirc[0.8ex] & \halfcirc[0.8ex] & \emptycirc[0.8ex] & \emptycirc[0.8ex] & \fullcirc[0.8ex] \tabularnewline
  API & Available calendars & \path{alexa.amazon.de/api/3PAccounts/accounts?includeLegacyCalendarAccounts=true}\colorcirc{pcb_purple}\colorcirc{pcb_red} & \halfcirc[0.8ex] & \emptycirc[0.8ex] & \emptycirc[0.8ex] & \emptycirc[0.8ex] & \emptycirc[0.8ex] & \fullcirc[0.8ex] \tabularnewline
  API & Calendar events & \path{alexa.amazon.de/api/calendar/events/getEvents/}\textrm{\tiny \textit{\{directedId\}}}\path{?startDateTime=}\textrm{\tiny \emph{\{ts\}}}\path{&endDateTime=}\textrm{\tiny \emph{\{ts\}}}\colorcirc{pcb_purple} & \halfcirc[0.8ex] & \emptycirc[0.8ex] & \emptycirc[0.8ex] & \emptycirc[0.8ex] & \emptycirc[0.8ex] & \fullcirc[0.8ex] \tabularnewline
  API & List of local users (with IDs) & \textrm{\tiny \textit{\{comA\}}}\path{/accounts/}\textrm{\tiny \textit{(\{directedId\})}}\colorcirc{pcb_red}\colorcirc{pcb_orange}\colorcirc{pcb_blue} & \emptycirc[0.8ex] & \emptycirc[0.8ex] & \emptycirc[0.8ex] & \emptycirc[0.8ex] & \emptycirc[0.8ex] & \fullcirc[0.8ex] \tabularnewline
  API & Contacts & \textrm{\tiny \textit{\{comA\}}}\path{/users/}\textrm{\tiny \textit{\{commsId\}}}\path{/paginatedContacts}\colorcirc{pcb_orange}\colorcirc{pcb_yellow} & \emptycirc[0.8ex] & \halfcirc[0.8ex] & \halfcirc[0.8ex] & \halfcirc[0.8ex] & \halfcirc[0.8ex] & \fullcirc[0.8ex] \tabularnewline
  API & List of conversations & \textrm{\tiny \textit{\{comA\}}}\path{/users/}\textrm{\tiny \textit{\{commsId\}}}\path{/conversations} & \emptycirc[0.8ex] & \fullcirc[0.8ex] & \emptycirc[0.8ex] & \fullcirc[0.8ex] & \fullcirc[0.8ex] & \fullcirc[0.8ex] \tabularnewline
  API & Conversation messages & \textrm{\tiny \textit{\{comA\}}}\path{/users/}\textrm{\tiny \textit{\{commsId\}}}\path{/conversations/}\textrm{\tiny \textit{\{conversationId\}}}\path{/messages}\colorcirc{pcb_orange} & \emptycirc[0.8ex] & \fullcirc[0.8ex] & \emptycirc[0.8ex] & \fullcirc[0.8ex] & \fullcirc[0.8ex] & \fullcirc[0.8ex] \tabularnewline
  API & Recent communication & \textrm{\tiny \textit{\{comA\}}}\path{/contacts/users/}\textrm{\tiny \textit{\{commsId\}}}\path{/recentCommunications}\colorcirc{pcb_orange}\colorcirc{pcb_yellow} & \emptycirc[0.8ex] & \emptycirc[0.8ex] & \emptycirc[0.8ex] & \emptycirc[0.8ex] & \emptycirc[0.8ex] & \fullcirc[0.8ex] \tabularnewline
  API & Device inform. \& communic. features & \textrm{\tiny \textit{\{comA\}}}\path{/homegroups/}\textrm{\tiny \textit{\{homegroupId\}}}\path{/devices}\colorcirc{pcb_orange} & \emptycirc[0.8ex] & \emptycirc[0.8ex] & \emptycirc[0.8ex] & \emptycirc[0.8ex] & \emptycirc[0.8ex] & \fullcirc[0.8ex] \tabularnewline
  API & Enabled skills & \path{skills-store.amazon.de/app/front-page/yourskills} & \halfcirc[0.8ex] & \emptycirc[0.8ex] & \emptycirc[0.8ex] & \emptycirc[0.8ex] & \emptycirc[0.8ex] & \fullcirc[0.8ex] \tabularnewline
  API & Photo \& video metadata & \path{cdws.eu-west-1.amazonaws.com/drive/v1/search}\colorcirc{pcb_purple} & \emptycirc[0.8ex] & \halfcirc[0.8ex] & \emptycirc[0.8ex] & \emptycirc[0.8ex] & \emptycirc[0.8ex] & \fullcirc[0.8ex] \tabularnewline
  API & Download photos \& videos & \path{content-eu.drive.amazonaws.com/v2/download/signed/}\textrm{\tiny \textit{\{photoId\}}}\path{/content?ownerId=}\textrm{\tiny \textit{\{ownerId\}}} & \emptycirc[0.8ex] & \halfcirc[0.8ex] & \emptycirc[0.8ex] & \emptycirc[0.8ex] & \emptycirc[0.8ex] & \fullcirc[0.8ex] \tabularnewline
  API & GraphQL; \emph{similar to \path{/api/phoenix}} & \path{alexa.amazon.de/nexus/v1/graphql} & \emptycirc[0.8ex] & \emptycirc[0.8ex] & \emptycirc[0.8ex] & \emptycirc[0.8ex] & \emptycirc[0.8ex] & \fullcirc[0.8ex] \tabularnewline
  \bottomrule
\end{tabular}

\begin{tablenotes}
  \item {[1]: \cite{chung2017alexaecosystem}},
    {[2]: \cite{youn2021echoshow}},
    {[3]: \cite{krueger2020using}},
    {[4]: \cite{olufohunsi:2020:alexa}},
    {[5]: \cite{hyde2017alexa}}
  \item Abbrev.: \textrm{\tiny \textit{\{AApp\}}} = \path{/data/data/com.amazon.dee.app/}, \textrm{\tiny \textit{\{PApp\}}} = \path{/data/data/com.amazon.clouddrive.photos/}, \textrm{\tiny \textit{\{comA\}}} = \path{alexa-comms-mobile-service.amazon.com}, 
  \emph{\{ts\}}= timestamp
  \item Kind of user ID revealed by artifact: \colorcirc{pcb_purple} = \textit{customerId}, \colorcirc{pcb_red} = \textit{directedId}, \colorcirc{pcb_orange} = \textit{commsId}, \colorcirc{pcb_yellow} = \textit{contactId}, \colorcirc{pcb_green} = \textit{personId}, \colorcirc{pcb_blue} = \textit{personIdV2}
  \item Coverage of artifact(s): \fullcirc[0.8ex] = artifact found,
    \halfcirc[0.8ex] = found similarly at another location,
    \emptycirc[0.8ex] = not found,
    ``\textbf{--}'' = appears to be deprecated

  \end{tablenotes}
\end{threeparttable}
}
\end{table*}

\subsection{Limitations and Future Work}
\label{discussion:limitations-future-work}

After our test data generation, we discovered that local logs on the Echo Show~15 
are deleted after about three days, which resulted in certain performed actions 
not being detectable in log files. %
To strive for a more systematic mapping of performed actions 
during test data generation (cf. \Cref{filesystem-artifacts:test-data-generation})
and corresponding persisted artifacts on the file system of the Echo Show~15, 
differential forensic analysis \citep{garfinkel:2012:differential:DI} 
could be performed based on more frequent and atomic file system images, as, for example, 
done by \citet{eichhorn2024steamdeck}. %
Additional usage scenarios could be considered during test data generation 
to take further user interactions and behaviors into account.

The failed cross-site request forgery check 
that occurred while testing the GraphQL \ac{API}  
(cf. \Cref{cloud-app:cloud:api-artifacts:graphql}) 
could be solved by analyzing the code 
of ``\path{com.amazon.alexa.identity.apk}'' found on the Echo Show~15,
albeit the legal situation for researchers in this regard may vary by country.

We identified unused pins for a Micro-HDMI port 
(cf. \Cref{hardware-examination:pcb:hdmi}) 
which we soldered to an HDMI breakout board
to test the Echo Show~15 as both input and output device. %
Although all soldered connections were checked for continuity, 
our attempts failed, leaving it for future work to figure out 
what the purpose of the Micro-HDMI port is, 
whether it needs to be activated in Fire OS, 
or what we have missed.

The database ``\path{map_data_storage_v2.db}'' (cf. \Cref{filesystem-artifacts:token-database}), 
was also used by a dedicated app for the fitness tracker \emph{Amazon Halo} 
\citep{hutchinson:2022:wearables}, which raises the question 
whether other smart devices by Amazon use the database as well, 
and whether the refresh tokens have the same scope of permission on all devices. %
Additionally, our database contained another token (\texttt{adptoken})
which is likely for the \ac{API} of Audible, Amazon's audiobook and podcast service, 
as well as further tokens with sensitive names 
whose particular purposes and contexts of use, however, 
have not been investigated yet 
(e.g., \texttt{privatekey}, \texttt{encrypt.key}).

At the time of writing, it became public that Amazon will 
discontinue to build their \acl{OS} upon Android
\citep{roettgers:2023:vega-os, verge:2023:vega-os-echo5}. 
While it is yet unknown
which devices will run the new \acs{OS} with the internal name \textit{Vega}, 
the unveiled pinout for the eMMC interface 
on the Echo Show 15 (cf. \Cref{hardware-examination:pcb:emmc}) 
could potentially become handy if Vega is rolled out on older devices 
and its file system remains unencrypted. %
Moreover, as the pinout for the \ac{eMMC} interface 
can be used for write access, it remains to be seen whether 
manipulations of the firmware become feasible, 
despite it being signed (cf. \Cref{hardware-examination:pcb:uart}). %

\section{Conclusion}
\label{conclusion}

In this paper, we forensically examined Amazon's smart display 
\emph{Echo Show 15} for the first time. %
On the \acl{MLB}, we discovered a working but restricted \ac{UART} port 
as well as undocumented \aclp{TP} for the \acs{eMMC} interface, 
resulting in non-invasive access to the unencrypted file system contents of \emph{FireOS}. %
Locally, we identified artifacts of forensic relevance 
about users' presence and usage behavior, 
including logged events of \emph{Visual ID} about 
movements and users detected by the built-in camera. %
Moreover, by trivially decrypting a ``refresh token'' 
which was stored jointly with the corresponding encryption key 
in the same database file, 
we were able to request a new ``access token'' 
for the Alexa \ac{API}, 
granting us access to user-related remote artifacts stored in Amazon's cloud, 
including Alexa voice requests, calendars, contacts, conversations, photos, and videos. %
Finally, we analyzed the network traffic of two companion apps,
namely \emph{Alexa} and \emph{Photos}, and identified new Alexa \ac{API} endpoints.
In terms of practical relevance, our findings show  
how forensic investigators can escalate their data acquisition procedure 
from local artifacts on a seized Echo Show 15 device 
to remote artifacts stored in the vendor's cloud 
without needing the suspect's or Amazon's cooperation 
for bypassing any security mechanism. %

\section*{Acknowledgments}
\label{acknowledgments}

We thank Felix Freiling for his precious support 
as well as the anonymous reviewers for their valuable comments. %
This work has been supported %
by the Bavarian Ministry of Science and Arts %
as part of the project ``Security in Everyday Digitization'' (ForDaySec). %


\bibliographystyle{cas-model2-names}
\bibliography{references}

\printcredits

\end{document}